\documentclass[mathpazo]{cicp}

\usepackage{graphicx}% Include figure files
\graphicspath{{Figures/}}
\usepackage{dcolumn}% Align table columns on decimal point
\usepackage{amsmath,bm} % bold math
\usepackage{amsbsy}
\usepackage[usenames,dvipsnames]{xcolor}
\usepackage{multirow} % Required for multirows
\usepackage{makecell}

\usepackage{color}
\definecolor{darkblue}{rgb}{0,0,0.6}
\definecolor{darkred}{rgb}{0.6,0,0}
\definecolor{darkgreen}{rgb}{0,0.6,0}
\usepackage[colorlinks=true,urlcolor=darkblue,citecolor=darkblue,linkcolor=darkred,hyperfootnotes=false]{hyperref}

\begin{document}
%%%%% title : short title may not be used but TITLE is required.
% \title{TITLE}
% \title[short title]{TITLE}
\title{Deep Learning-Based Computational Method for Soft Matter Dynamics: Deep Onsager-Machlup Method}

%%%%% author(s) :
% single author:
% \author[name in running head]{AUTHOR\corrauth}
% [name in running head] is NOT OPTIONAL, it is a MUST.
% Use \corrauth to indicate the corresponding author.
% Use \email to provide email address of author.
% \footnote and \thanks are not used in the heading section.
% Another acknowlegments/support of grants, state in Acknowledgments section
% \section*{Acknowledgments}
% \author[O.~Author]{Only Author\corrauth}
% \address{School of Mathematical Sciences, Beijing Normal University,
% Beijing 100875, P.R. China}
% \email{{\tt author@email} (O.~Author)}

\author[Li Z H et.~al.]{Zhihao Li\affil{1,2}$^{\dag}$,
      Boyi Zou\affil{3,4}$^{\dag}$, Haiqin Wang\affil{5,2}, Jian Su\affil{5,2}, Dong Wang\affil{3,4*} and Xinpeng Xu\affil{2,5}\corrauth}
\address{\affilnum{1}\ Department of Civil Engineering and Smart Cities, College of Engineering, Shantou University, Shantou, 515063, Guangdong, China. \\
         \affilnum{2}\ Department of Physics and MATEC Key Lab, Guangdong Technion - Israel Institute of Technology, 241 Daxue Road, Shantou, 515063, Guangdong, China. \\
         \affilnum{3}\ School of Science and Engineering, The Chinese University of Hong Kong, Shenzhen, 518172 Guangdong, China. \\
         \affilnum{4}\ Shenzhen International Center for Industrial and Applied Mathematics, Shenzhen Research Institute of Big Data, Shenzhen, 518172, Guangdong, China. \\
         \affilnum{5}\ Technion – Israel Institute of Technology, Haifa, 32000, Israel. \\
         }
\emails{{\tt xu.xinpeng@gtiit.edu.cn} (X. Xu), {\tt wangdong@cuhk.edu.cn} (D. Wang), {\tt zhihao.li@gtiit.edu.cn} (Z. Li), {\tt  221019226@link.cuhk.edu.cn} (B. Zou), {\tt cloverwhq@gmail.com} (H. Wang), {\tt jian.su@gtiit.edu.cn} (J. Su)}

\footnotetext{\textit{$^{\dag}$~These authors contributed equally to this work.}}

% \footnote and \thanks are not used in the heading section.
% Another acknowlegments/support of grants, state in Acknowledgments section
% \section*{Acknowledgments}

% multiple authors:
% Note the use of \affil and \affilnum to link names and addresses.
% The author for correspondence is marked by \corrauth.
% use \emails to provide email addresses of authors
% e.g. below example has 3 authors, first author is also the corresponding
%      author, author 1 and 3 having the same address.
% \author[Zhang Z R et.~al.]{Zhengru Zhang\affil{1}\comma\corrauth,
%       Author Chan\affil{2}, and Author Zhao\affil{1}}
% \address{\affilnum{1}\ School of Mathematical Sciences,
%          Beijing Normal University,
%          Beijing 100875, P.R. China. \\
%           \affilnum{2}\ Department of Mathematics,
%           Hong Kong Baptist University, Hong Kong SAR}
% \emails{{\tt zhang@email} (Z.~Zhang), {\tt chan@email} (A.~Chan),
%          {\tt zhao@email} (A.~Zhao)}
% \footnote and \thanks are not used in the heading section.
% Another acknowlegments/support of grants, state in Acknowledgments section
% \section*{Acknowledgments}

%%%%% Begin Abstract %%%%%%%%%%%
\begin{abstract}
A deep learning-based computational method is proposed for soft matter dynamics -- the deep Onsager-Machlup method (DOMM). It combines the brute forces of deep neural networks (DNNs) with the fundamental physics principle -- Onsager-Machlup variational principle (OMVP). In the DOMM, the trial solution to the dynamics is constructed by DNNs that allow us to explore a rich and complex set of admissible functions. It outperforms the Ritz-type variational method where one has to impose carefully-chosen trial functions. This capability endows the DOMM with the potential to solve rather complex problems in soft matter dynamics that involve multiple physics with multiple slow variables, multiple scales, and multiple dissipative processes. Actually, the DOMM can be regarded as an extension of the deep Ritz method (DRM) developed by E and Yu that uses DNNs to solve static problems in physics. In this work, as the first step, we focus on the validation of the DOMM as a useful computational method by using it to solve several typical soft matter dynamic problems: particle diffusion in dilute solutions, and two-phase dynamics with and without hydrodynamics. The predicted results agree very well with the analytical solution or numerical solution from traditional computational methods. These results show the accuracy and convergence of DOMM and justify it as an alternative computational method for solving soft matter dynamics.
\end{abstract}
%%%%% end %%%%%%%%%%%

%%%%% AMS/PACs/Keywords %%%%%%%%%%%
%\pac{}

%%%% maketitle %%%%%
\maketitle

%%%% Start %%%%%%

\section{Introduction} \label{Sec:Intro}

Approximate variational methods such as the Ritz-type method have been widely used to study the structure, phase behaviors, and dynamics of soft matter~\cite{Sam2018,Doi2013,Doi2019,Reddy2017}. These methods are mostly based on the following variational principles: the variational principle of minimum free energy (MFEVP)~\cite{Sam2018,Doi2013,Reddy2017}, Onsager’s variational principle (OVP)~\cite{Onsager1931a,Onsager1931b,Doi2021}, and the Onsager-Machlup variational principle (OMVP)~\cite{Onsager1953,Doi2019}. In these variational methods, some trial functions to the problem are assumed empirically where the state variables are taken as combinations of some analytical functions with a much smaller number of adjustable parameters~\cite{Reddy2017,Doi2021,Xu2021,Xu2022}. Such methods simplify the problem significantly: they bypass the derivation and solution of the complicated governing differential equations and go directly from the variational statement to an approximate solution to the problem. 

Recently, it has been proposed that the power of these variational methods can be further enhanced if we use deep neural networks (DNNs) to construct the trial functions. For example, the deep Ritz method (DRM)~\cite{Weinan2018} uses variational formulations of physics models (for example from MFEVP) to solve static problems in physics. In our former work~\cite{Xu2022}, we have applied DRM to the spontaneous bending of active elastic solids. Actually, DRM is regarded as one of the DNN methods for solving partial differential equations (PDEs). Among other DNN methods, the most popular methods are the Physics-Informed Neural Network (PINN)~\cite{Perdikaris2019} and its variants~\cite{Karniadakis2021physics} where the residual of a PDE is minimized as a loss function. Another type of DNNs for the approximate solutions of PDEs is the data-driven approach. It involves collecting a large amount of data under similar conditions for the same PDE to learn a solution operator that directly maps the conditions of the PDE to its solution. Examples of such methods include Deep Operator Neural Network (DeepONet)~\cite{lu2021learning}, Fourier Neural Operator (FNO)~\cite{li2020fourier}, OnsagerNet~\cite{yu2021onsagernet}, Variational Onsager Neural Networks (VONNs)~\cite{Reina2021VONN}, flow maps~\cite{qin2019data}, and many others. The main difference between the data-driven approach and the mechanism-driven approach is that the mechanism-driven does not require data and can be treated similarly to those in traditional methods. All these deep learning methods have recently been used in many different problems in soft matter physics, prediction of microphase-separated structures in diblock colpolymers~\cite{Chen2018}, stochastic control of colloidal self-assembly~\cite{Nodozi2023PINN}, phase separation dynamics~\cite{zhao2021PINN}, bubble growth dynamics~\cite{Karniadakis2021PINN}. For more discussions and applications about DNNs for solving PDEs, we recommend the review paper by Karniadakis \emph{et al}~\cite{Karniadakis2021physics}.

Here we are particularly interested in deep-learning methods for solving PDEs such as DRM that combine variational principles with deep learning. Recently several different such methods have been proposed. For example, Variational Physics Informed Neural Network (VPINN)~\cite{Karniadakis2019variational} uses the weak form of partial differential equations (PDEs) to construct the loss function. Weak Adversarial Network (WAN)~\cite{Zang_2020, bao2024wanco} further utilizes an adversarial network to approximate the trial functions via formulating the original problems into minimax problems. Energetic Variational Neural Network (EVNN)~\cite{ChunLiu2022} proposes a temporal-then-spatial approach to solving $L^2$-gradient flows and generalized diffusion by using the minimizing movement. These deep learning methods based on variational principles improve the training accuracy and enable the trained solutions to maintain certain properties, such as energy dissipation. They are mostly based on the universal approximation theorem proposed about thirty years ago stating that DNNs could approximate continuous functions arbitrarily close~\cite{cybenko1989approximation,Chen1993} and fit PDE solutions~\cite{lagaris1998artificial}. They basically use DNNs to represent the solution and transform PDEs into different forms (such as strong, weak, and variational forms) to construct the loss functions required for the training of DNNs. In this work, we follow a similar idea and propose a unified computational method for soft matter dynamics -- the deep Onsager-Machlup method (DOMM), which combines the brute forces of DNNs with the fundamental physics principle, OMVP. We then apply the DOMM to solve some typical dynamic problems in soft matter such as diffusion in dilute solutions and two-phase dynamics with and without hydrodynamics. 

The rest of the paper is organized as follows. In Section~\ref{Sec:DOMM}, we provide a brief review of the Onsager-Machlup variational principle (OMVP) and introduce the deep Onsager-Machlup method (DOMM).  In Secs.~\ref{Sec:App1}-\ref{Sec:App3}, we then apply the DOMM to solve several typical dynamic problems in soft matter physics: particle diffusion in dilute solutions, two-phase dynamics with and without hydrodynamics, respectively. The results predicted by the DOMM agree very well with known solutions, which validate the DOMM and justify its accuracy and convergence as an alternative computational method in solving soft matter dynamics. Finally, in the section~\ref{Sec:Conclusions}, we summarize our major results and make some general remarks.

\section{Deep Onsager-Machlup method (DOMM) for soft matter dynamics}~\label{Sec:DOMM}

\subsection{Onsager-Machlup variational principle (OMVP)}

In a soft matter system described by a set of slow state variables $\boldsymbol{\alpha}=\left(\alpha_{1}, \alpha_{2}, \ldots \right)$, we consider its dynamic evolution from some initial state $\boldsymbol{\alpha}_0$ at time $t_0$ to the final state $\boldsymbol{\alpha}$ at time $t$. In the Onsager-Machlup theory for linear irreversible processes~\cite{Onsager1953,Wang2019RMP,Xu2021}, the transition probability between the two states is given by
\begin{align}\label{eq:DOMM-Plong}
P\left(\boldsymbol{\alpha}, t \mid \boldsymbol{\alpha}_0,t_0\right) \propto \int {\mathcal D} \boldsymbol{\alpha}(t) \exp\left\{-\frac{\mathcal{L}_{\mathrm{om}}[\boldsymbol{\alpha}(t)]}{2k_{\mathrm B}T}\right\},
\end{align} 
with $T$ being the temperature and $k_{\mathrm B}$ being the Boltzmann constant. Here $\mathcal{L}_{\mathrm{om}}[\boldsymbol{\alpha}(t)]$ is the Onsager-Machlup (OM) functional or integral, taking the following form~\cite{Onsager1953,Graham1977,Weinan2004,Ren2020,Doi2019,Xu2021}
\begin{equation}\label{eq:OVP-Olong-LeastSquare}
\mathcal{L}_{\mathrm{om}}\equiv \int_{t_0}^{t} dt'\left[\sum_{i,j} \frac{1}{2}\zeta_{ij}(\dot{\alpha}_i- \dot{\alpha}_i^*)(\dot{\alpha}_j-\dot{\alpha}_j^*) +\sum_{i} \frac{1}{2} {\partial_{\alpha_i} f_i}\right],
\end{equation} 
with $\zeta_{ij}$ being the ''constant" resistance (or friction) coefficient matrix, $\dot{\boldsymbol{\alpha}}$ being the time derivative of the state variables ${\boldsymbol{\alpha}}$, $\dot{\alpha}_i^*(\boldsymbol{\alpha},t) = \mu_{ik} f_k(\boldsymbol{\alpha},t)$ being the actual rates of the system at the state $\boldsymbol{\alpha}$ and time $t$, $\mu_{ik}=\zeta^{-1}_{ik}$ being the ''constant" mobility coefficient matrix. The general force $\mathbf{f}$ includes two parts, $\mathbf{f}=\mathbf{f}_{\mathrm{c}} +\mathbf{f}_{\mathrm{a}}$, with $\mathbf{f}_{\mathrm{c}} = -\partial_{\boldsymbol{\alpha}} {\mathcal F}$ being the conservative force derived from the restricted free energy ${\mathcal F}(\boldsymbol{\alpha},t)$ and the non-conservative active force $\mathbf{f}_{\mathrm{a}}$ that cannot be derived from any energy function~\cite{Wang2019RMP,Xu2021,Xu2022}. Hereafter, we use the subscript such as $\partial_{\boldsymbol{\alpha}}$ to denote partial derivatives $\partial/\partial{\boldsymbol{\alpha}}$. 

The \emph{Onsager-Machlup variational principle} (OMVP)~\cite{Onsager1953,Weinan2004,Ren2020,Doi2019,Xu2021} states that nature tends to choose the most probable kinetic path that minimizes the OM functional $\mathcal{L}_{\mathrm{om}}[\boldsymbol{\alpha}(t)]$ with respect to the function $\boldsymbol{{\alpha}}(t)$. The OMVP is a global principle that can determine the most probable path taking the system to the far future $t\gg t_0$ under various constraints. For example, it can be used to determine the long-time behaviors such as the steady-states~\cite{Doi2019,Doi2021} and to locate the optimal transition pathway (so-called minimal action path, MAP) of rare events that takes the system from one local free-energy minimum to the global free-energy minimum~\cite{Weinan2004,Touchette2009,Ren2020}. In comparison, if we only want to find the most probable state in the immediate future (for example, the next infinitesimal time-step, without additional constraints to the final states), we should use the local version of the OMVP -- \emph{Onsager's variational principle} (OVP)~\cite{Onsager1931a,Onsager1931b,Onsager1953,Xu2021,Doi2021}, which states that nature chooses the kinetic path that minimizes the OM functional $\mathcal{L}_{\mathrm{om}}[\boldsymbol{\alpha}(t)]$ with respect to the rate of the change of state variables $\boldsymbol{\dot{\alpha}}(t)$. This is exactly the case in all the studies carried out in this work. In this case, since the last term in Eq.~(\ref{eq:OVP-Olong-LeastSquare}) depends only on the state variable $\boldsymbol{\alpha}$ and time $t$, when we apply OVP to predict the temporal evolution of the system, we only need to minimize the first (non-negative) quadratic integrand term of Eq.~(\ref{eq:OVP-Olong-LeastSquare}), which equals to zero (the minimum) only when $\dot{\boldsymbol{\alpha}}$ takes the value on the actual kinetic path, {\emph{i.e.}, $\dot{\boldsymbol{\alpha}}=\dot{\boldsymbol{\alpha}}^*$}.

Variational principles like MFEVP, OMVP, and OVP provide not only an equivalent substitute for the applications of kinetic (force balance) equations but also some powerful direct variational methods of finding approximate solutions to these equations, \emph{e.g.}, Ritz method and the least-squares method~\cite{Reddy2017,Xu2021,Xu2022}. In these approximate variational methods, some trial (admissible) solutions to the equations are assumed where the state variables $\boldsymbol{\alpha}$ are taken as combinations of some simple analytical functions with a much smaller number of adjustable parameters, $\mathbf{c}$. Particularly, in soft matter physics, such variational methods, usually referred to as Ritz-type methods~\cite{Xu2022}, have been widely used to solve stationary problems such as the structure and phase behaviors of soft matter~\cite{Doi2013,Xu2022}. Recently, Doi~\cite{Doi2015} followed a similar idea of the Ritz-type method and used the OMVP and OVP as an approximate approach to soft matter dynamics~\cite{Doi2015,Doi2019,Doi2021}.
 
More specifically, in the continuum field theory of linear irreversible processes~\cite{Doi2015,Xu2021}, Doi's approximate variational method can be summarized as follows. The state (field) variables are denoted by $\boldsymbol{\alpha}(\mathbf{x},t)$ with $\mathbf{x}$ being the spatial position and the OM integral is a functional of $\boldsymbol{\alpha}(\mathbf{x},t)$, \emph{i.e.}, ${\mathcal L}_{\mathrm{om}}={\mathcal L}_{\mathrm{om}}[\boldsymbol{\alpha}(\mathbf{x},t)]$. One usually seeks an explicit approximation (trial solution) $\mathbf{A}_{\mathrm N}(\mathbf{x},t)$ to $\boldsymbol{\alpha}(\mathbf{x},t)$, for example, in the finite series form of~\cite{Xu2021,Xu2022} 
\begin{equation}\label{Eq:DOMM-OMVP-Trial}
\boldsymbol{\alpha}(\mathbf{x},t)\approx \mathbf{A}_{\mathrm N}(\mathbf{x};\mathbf{c}(t)) =\sum_{i=1}^{N} c_{i}(t) \boldsymbol{\phi}_{i}(\mathbf{x}),
\end{equation}
% \begin{equation}\label{Eq:DOMM-OMVP-Trial}
% \boldsymbol{\alpha}(\mathbf{x},t)\approx \mathbf{A}_{\mathrm N}(\mathbf{x};\mathbf{c}(t)) =\sum_{i=1}^{N} c_{i}(t) \mathbf{\phi}_{i}(\mathbf{x}),
% \end{equation}
in which $\boldsymbol{\phi}_{i}(\mathbf{x})$ are basis functions, $c_{i}(t)$ are the unknown time-dependent parameters, and $N$ is a fixed and pre-selected number of parameters $c_{i}(t)$. 

Substituting the trial solution $\mathbf{A}_{\mathrm N}$ into the OM integral $\mathcal{L}_{\mathrm{om}}$, we then obtain $\mathcal{L}_{\mathrm{om}}$ as a function of the parameters $\mathbf{c}(t)$ and their time derivatives: $\mathcal{L}_{\mathrm{om}}(\dot{\mathbf c}, \mathbf{c})$. Minimizing it with respect to the rates $\mathbf{\dot{c}}$ gives a set of ordinary differential equations of $\mathbf{c}(t)$, and its solution (substituted back into Eq.~(\ref{Eq:DOMM-OMVP-Trial})) gives $\mathbf{A}_{\mathrm N}(\mathbf{x},t)$, providing an approximate solution to the system dynamics. Such direct variational methods simplify the analysis of the system dynamics significantly~\cite{Reddy2017,Doi2019,Doi2021,Xu2021,Xu2022}: they bypass the derivation and solution of the complex governing dynamic equations in the original high-dimensional $\boldsymbol{\alpha}$--space, and go directly from the variational statement to an approximate solution of the problem in a much lower-dimensional $\mathbf{c}$--space. 

\begin{figure}[ht]
\centering
\includegraphics[width = 0.5\linewidth]{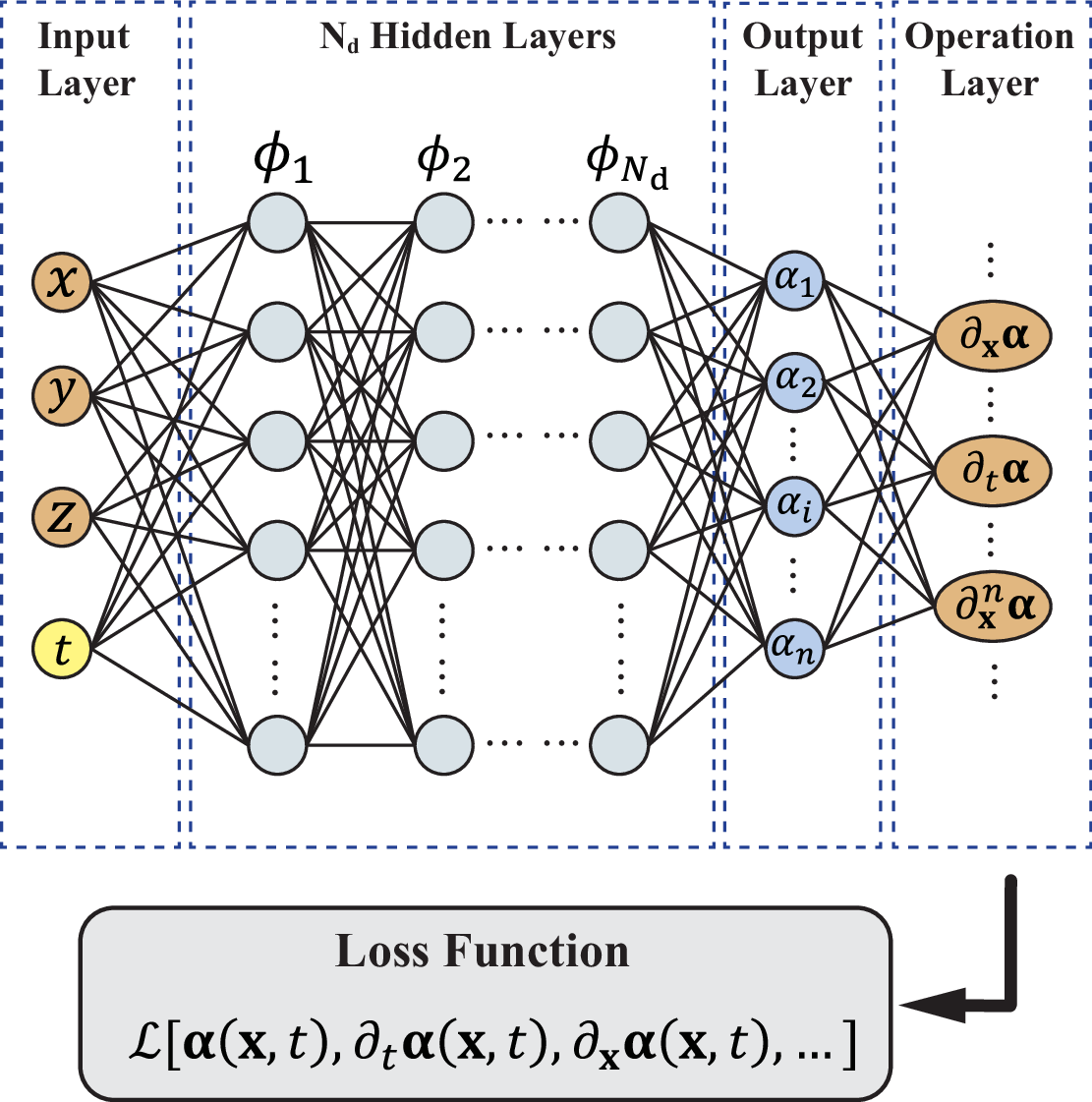}
\caption{Schematic illustration of the deep Onsager-Machlup method based on deep neural networks. The structure of the deep neural network with a depth $N_{\mathrm{d}}$ (\emph{i.e.}, consisting of $N_{\mathrm{d}}$ hidden layers) and width $N_\mathrm{w}$ is shown, where the hidden layers are connected by one linear transformation, activation, and a residual connection as explained in Eq.~(\ref{Eq:DOMM-MLNet}). The input of the neural network is the temporal node $t$ and the three-dimension (3D) spatial coordinates $\mathbf{x}=\left\{x,y,z\right\}$. The output is the solution of a $n$-dimension set of slow variables $\boldsymbol{\alpha}(\mathbf{x},t)=\{\alpha_1, \ldots, \alpha_n\}$.
In the last operation layer, the corresponding (spatial and/or temporal) derivatives are computed through the chain rule of automatic differentiation by {\tt autograd} in {\tt pytorch}. 
} \label{Fig:DOMM_Networks}
\end{figure}

\subsection{DOMM: A deep learning-based method of solving variational dynamics}

Recently, following the idea of the Ritz-type method, a "deep Ritz method" (DRM)~\cite{Weinan2018} based on deep learning and deep neural networks (DNNs) has been proposed to numerically solve stationary problems in physics~\cite{Weinan2018}. In the DRM, neural networks are used to construct the trial solutions, and the free energy functional ${\mathcal F}$ is minimized as a major loss function during the ``training" phase of the network by varying the undetermined network parameters. The representation of the trial solution using DNNs allows us to explore a rich and complex set of admissible trial functions. The trained neural network is a map between a sampling (observation) dataset of spatial coordinates (network input, $\{\mathbf{x}_i \}$) and the components of the state variable functions (network output, $\boldsymbol{\alpha}=\{\alpha_1, \alpha_2, \ldots, \alpha_n \}$) that minimizes the free energy as a loss function. 

In this work, we combine the idea of the DRM with the approximate variational method proposed by Doi group~\cite{Doi2015,Doi2019} and propose a ``deep Onsager-Machlup method" (DOMM) to solve problems in soft matter dynamics. 
Firstly, as in the DRM, the trial solution in the DOMM is also constructed using a neural network (as shown schematically in Fig.~\ref{Fig:DOMM_Networks} in the composite-function form of 
\begin{equation}\label{Eq:DOMM-MLTrial}
\boldsymbol{\alpha}(\mathbf{x},t)\approx \mathbf{A}_{\mathrm N}(\mathbf{x},t; \mathbf{c})=\mathbf{a} \cdot [\boldsymbol{\phi}_{N_{\mathrm{d}}} \circ \ldots \circ \boldsymbol{\phi}_{1}(\mathbf{x},t;\tilde{c}_1)]+\mathbf{b},
\end{equation} 
in which the trial solution (network output) $\mathbf{A}_{\mathrm N}(\mathbf{x},t)\in \mathbb{R}^{n}$; $\boldsymbol{\phi}_{i}(\mathbf{x},t;\tilde{c}_i)$ represents the $i$-th hidden layer of the neural network, typically taking the form of
\begin{equation}\label{Eq:DOMM-MLNet}
\boldsymbol{\phi}_{i}(\mathbf{s};\tilde{\mathbf{c}}_i)= {\varphi}\left(\mathbf{W}_{i}  \mathbf{s} +\mathbf{b}_{i}\right)+\mathbf{s},
\end{equation}
with $\mathbf{s}\in \mathbb{R}^{N_\mathrm{w}}$ representing the sampling (observation) dataset of spatiotemporal coordinates (network input, $\mathbf{x}$ and $t$). Here the neural network has a depth $N_{\mathrm{d}}$ (\emph{i.e.}, consisting of $N_{\mathrm{d}}$ hidden layers) and width $N_\mathrm{w}$. The full set of undetermined parameters in the trial solution is $\mathbf{c}=\{\tilde{\mathbf{c}}, \mathbf{a}, \mathbf{b}\}$, in which $\mathbf{a} \in \mathbb{R}^{n\times N_\mathrm{w}}$ and $\mathbf{b} \in \mathbb{R}^{n}$ define a linear transformation that reduces the dimension of the network output to the desired smaller dimension $n$ of the trial functions $\mathbf{A}_{\mathrm N}$, and $\tilde{\mathbf{c}}=\{\tilde{\mathbf{c}}_i\}=\{\mathbf{W}_{i} \in \mathbb{R}^{N_\mathrm{w} \times N_\mathrm{w}}, \mathbf{b}_{i}  \in \mathbb{R}^{N_\mathrm{w}}\}$ denotes the parameters associated with the linear transformations in $i$-th hidden layer. The total number of parameters in each hidden layer is $N_\mathrm{w}^2+N_\mathrm{w}$, and hence the total number of undetermined network parameters or the dimension of $\mathbf{c}$ in the trial solution $\mathbf{A}_{\mathrm N}$ is $N=N_\mathrm{d}(N_\mathrm{w}^2+N_\mathrm{w})+n(N_\mathrm{w}+1)$.
$\varphi(\mathbf{s})$ is the (scalar) activation function whose smoothness plays a key role in the accuracy of the algorithm. Many different types of activation functions have been used~\cite{nwankpa2018activation} such as $\mathrm{Sigmoid}$, $\mathrm{Tanh}$, $\mathrm{SoftMax}$, $\mathrm{ReLU}$, and $\mathrm{ReLU}^3$. In this work, we choose $\mathrm{Tanh}^3$ for all the training processes based on the empirical observation during the network training.

The last term in Eq.~(\ref{Eq:DOMM-MLNet}) is the residual connection that makes the network much easier to be trained since it helps to avoid the problem of vanishing gradients and makes the network deeper and more stable. 
Note that both the input $\mathbf{s}$ and the output $\boldsymbol{\phi}_{i}(\mathbf{s};\tilde{\mathbf{c}}_i)$ of each network layer are vectors in $\mathbb{R}^{N_\mathrm{w}}$. However, the input $\mathbf{x}=\{x,y,z\}$ and $t$ for the first block is in $\mathbb{R}^{4}$, not $\mathbb{R}^{N_\mathrm{w}}$. To resolve this problem, we can either pad $\mathbf{x}$ and $t$ by a zero vector when $N_\mathrm{w}>4$, or apply a linear transformation on $\mathbf{x}$ and $t$ when $N_\mathrm{w}<4$. 

In the practical solutions of soft matter dynamics using DOMM, the loss functions include not only the OM integral but also include other contributions: 
\begin{equation}\label{Eq:DOMM-Loss} 
\mathcal{L}=w_{\mathrm{om}} \mathcal{L}_{\mathrm{om}}+w_{\mathrm{bc}}\mathcal{L}_{\mathrm{bc}}+w_{\mathrm{ic}}\mathcal{L}_{\mathrm{ic}}+w_{\mathrm{con}}\mathcal{L}_{\mathrm{con}}, 
\end{equation} 
in which $\mathcal{L}_{\mathrm{om}}$ is the OM integral that approaches zero (the minimum) as the trial solution converges to the true solution; $\mathcal{L}_{\mathrm{bc}}$, $\mathcal{L}_{\mathrm{ic}}$, and $\mathcal{L}_{\mathrm{con}}$ represent additional loss function terms associated with boundary condition, initial conditions, and physical or geometrical constraints, respectively. The expressions of these additional loss function terms are taken to be the same quadratic form as in PINN~\cite{Perdikaris2019}. Ideally, each term of the loss function in Eq.~(\ref{Eq:DOMM-Loss}) needs to trend towards zero as training progresses. Here $w_{\mathrm{om}}$, $w_{\mathrm{con}}$, $w_{\mathrm{ic}}$, and $w_{\mathrm{bc}}$ represent their weight coefficients in the loss functions, respectively, as hyper-parameters. 
Note that, for a given soft matter system, all four loss function terms in Eq.~(\ref{Eq:DOMM-Loss}) can be found out systematically based on the Onsager-Machlup variational principle. The only arbitrariness in the loss function $\mathcal{L}$ lies in the choice of weight coefficients (hyper-parameters) in each loss function term, which is, actually, a general problem in most deep-learning methods for solving PDEs. As in other methods, we also need to manually adjust these weight coefficients based on specific dynamic problems. However, on the other hand, for dynamic problems involving multiphysics couplings (multiple dynamics equations in couplings), DOMM has some unique advantages over PINN because the coefficients in each term of the OM integral $\mathcal{L}_{\mathrm{om}}$ has specific physical meanings and don’t need to adjust. Therefore, the number of adjustable weight coefficient hyper-parameters is less than that of PINN. In the following sections, we will show how to find and give specific expressions of the loss function for each specific dynamic problem.

The training process of the neural network is to minimize the total loss function $\mathcal{L}$ in Eq.~(\ref{Eq:DOMM-Loss}). From the output (the trial state functions) of the neural network, the spatiotemporal derivatives of the trial state functions in the loss functions are computed through the chain rule of automatic differentiation by {\tt autograd} in {\tt pytorch}. The integration over space and time is computed using the Monte Carlo method. The loss function is minimized iteratively by the stochastic gradient descent method (SGD), where the undetermined network parameters $\mathbf{c}$ are updated iteratively by
\begin{equation}\label{Eq:DOMM-SGD}
\mathbf{c}^{m+1}=\mathbf{c}^{m}-\frac{\eta_{\mathrm{r}}}{N_{\mathrm k }}\sum_{k=1}^{N_{\mathrm k}}\nabla_{\mathbf{c}} \mathcal{L}[\mathbf{A}_{\mathrm N}(\mathbf{x}_{i_k}^m,t_{i_k}^m; \mathbf{c}^{m})],
\end{equation}
where $\eta_{\mathrm{r}}$ is the learning rate, $\{\mathbf{x}_{i_k}^m,t_{i_k}^m\}_{k=1}^{N_\mathrm k}$ is a data subset of the full sampling data $\{\mathbf{x}_{i}^m,t_{i}^m\}_{i=1}^{N_\mathrm s}$ with ${N_\mathrm k}$ ($<{N_\mathrm s}$) being the prescribed dimension of the data subset and ${N_\mathrm s}$ being the dimension of full sampling dataset (\emph{i.e.}, the number of spatiotemporal coordinates). Here we have substituted the trial solution $\mathbf{A}_{\mathrm N}(\mathbf{x},t; \mathbf{c})$ in Eq.~(\ref{Eq:DOMM-MLTrial}) into the loss functions in Eq.~(\ref{Eq:DOMM-Loss}) and discretized it properly. The data subset is chosen randomly from the full sampling dataset in every iteration. The standard {\tt Adam}~\cite{kingma2014adam, Weinan2018} optimizer is used to accelerate the training of the neural network.

Note that in comparison to traditional computational methods such as the finite difference method (FDM), the finite element method (FEM), or the spectral method (SM), using DNNs to solve PDEs is known to have the following advantages. Firstly, it is a mesh-free method that generates training datasets randomly, subject to some properly chosen distributions~\cite{caflisch_1998}. 
Empirically, such numerical (mesh-free) treatment, particularly for the integration of the loss function over the spatiotemporal coordinate space, is believed to be able to avoid the curse of dimensionality (\emph{i.e.}, the amount of data needed grows exponentially with the dimensionality of input coordinates), the problem of being trapped into local minimum states so that they can efficiently and accurately handle complex domains (such as perforated domains), and the over-fitting problems that may occur when the energy functional is discretized by any fixed spatial-grid points (for example, the Trapezoidal discretization proposed in the deep energy method, DEM~\cite{nguyen2020deep}).

Moreover, the DNNs deal with derivatives directly using the chain rule instead of numerical differentiation. This reduces the complexity of solving higher-order PDE problems, whereas traditional computational methods require careful construction of specific schemes for the discretization of derivatives to avoid numerical issues, such as stability issues.

\section{Application 1: Solving particle diffusion dynamics}\label{Sec:App1}

\subsection{Solving 1D diffusion dynamics}\label{Sec:App1_1}

As the first application of DOMM, we consider the one-dimensional (1D) diffusion of particles in dilute solutions~\cite{Doi2013,Doi2015,Doi2019}. In this case, the state variable is $n(x,t)$, the number density of particles at position $x$ and time $t$. For solutions where there is no process (such as chemical reactions) creating or annihilating the particles, the particle density $n$ can change only by diffusion, satisfying the conservation equation 
\begin{equation}\label{Eq:App1-nt}
\partial_t{n}+{\partial_x j}=0,
\end{equation}
where $j(x,t)$ is the flux of the particles. The free energy of the system is given by 
\begin{equation}\label{Eq:App1-F}
\mathcal{F} [n(x,t)]=\int_{-L}^{L} d x nk_{\mathrm{B}} T \ln (nv_0),
\end{equation}
with $v_0$ being the molecular volume and $L$ being the half-length of the system. The dissipation function $\Phi$ takes the following quadratic form
\begin{equation}\label{Eq:App1-Phi}
\Phi[j(x,t)]=\int_{-L}^L d x\frac{\zeta}{2n} j^{2},
\end{equation}
with $\zeta$ being the friction constant of one particle. Minimizing the Rayleighian $\mathcal{R}[j(x,t)]=\dot{\mathcal{F}}+\Phi$ with respect to the flux $j$ yields the most probable flux 
\begin{equation}\label{Eq:App1-jstar}
j^{*}=- D\partial_x n,
\end{equation}
with dot superscript denoting the time derivative and $D={k_{\mathrm{B}} T}/{\zeta}$ being the diffusion constant. Substituting Eq.~(\ref{Eq:App1-jstar}) into the conservation equation (\ref{Eq:App1-nt}) gives the standard diffusion equation
\begin{equation}\label{Eq:App1-1DEqn}
{\partial_t n}-D{\partial_x^{2} n}=0,
\end{equation}
and the Onsager-Machlup integral is given by
\begin{equation}\label{Eq:App1-OM}
% \mathcal{L}_{\mathrm{om}} = \int_0^{t}d{t} \left\{\mathcal{R}[j]-\mathcal{R}[j^*]\right\}= 
\mathcal{L}_{\mathrm{om}}=\int_0^{t}d{t}\int_{-L}^L d{x}\frac{\zeta}{2n}\left({j}-{j^*}\right)^2. 
\end{equation}

To be specific, we use the DOMM to solve the 1D diffusion dynamics for $-L\leqslant x \leqslant L$ and $t \geqslant 0 $ under two different sets of initial and boundary conditions as follows. 
\begin{itemize}
    \item Initial condition of $n_0=0.5+0.45 \,\mathrm{sgn}\left({x}\right)$ and the Neumann boundary condition of $\partial_x n(\pm L,t)=0$ with $\mathrm{sgn}(x)$ being the sign function, taking the value of $-1$ for $x<0$, $0$ for $x=0$, and $1$ for $x>0$. In this case, the analytical solution is 
\begin{equation}\label{Eq:App1-Sol1}
\begin{aligned}
n(x,t) &=0.5+\sum_{k=0}^{\infty}\frac{0.9}{(k+1/2)\pi}\sin\left[\left(k+{1}/{2}\right)\frac{\pi x}{L} \right]\times \exp{\left[-\left(k+{1}/{2}\right)^2 \frac{{\pi}^2t}{\tau_0}\right]},
\end{aligned}
\end{equation}
with $\tau_0\equiv L^2/D$ being the characteristic diffusion time. 

\item Initial condition of sinusoidal function sums ${n}_0=\sum_{k=1}^{4}\sin(k\pi {x}/L)$ and the Dirichlet boundary condition of $n(\pm L,t)=0$. In this case, the analytical solution is 
\begin{equation}\label{Eq:App1-Sol2}
n(x,t) = \sum_{k=1}^4 \exp\left({-k^2\pi^2t/\tau_0}\right) \sin(k\pi x/L).
\end{equation}
\end{itemize}
To use the DOMM to solve the above two 1D diffusion problems, the input parameters of the neural network are position $x$ and time $t$, while the output parameters are chosen to be the number density $n$ and the particle flux $j$. The loss function to be minimized to train the neural network takes the general form of Eq.~(\ref{Eq:DOMM-Loss}). 
Here the first loss function term $\mathcal{L}_{\mathrm{om}}$ is the Onsager-Machlup integral given in Eq.~(\ref{Eq:App1-OM}). The second loss function term is associated with boundary conditions: for case (i) with Neumann boundary condition, $\mathcal{L}_{\mathrm{bc}} = \int_0^{t}d{t'} \{[\partial_x n(-L,t')]^2+[\partial_x n(L,t')]^2\}$; for case (ii) with Dirichlet boundary condition, 
$\mathcal{L}_{\mathrm{bc}} = \int_0^{t}d{t'} [n^2(-L,t')+n^2(L,t')]$. The third loss function term is given by $\mathcal{L}_{\mathrm{ic}}=\int_{-L}^Ld{x}\left[n\left(x, 0\right)-n_0\right]^2$, which imposes the initial conditions. The last loss function term is given by $\mathcal{L}_{\mathrm{con}}=\int_0^{t}d{t'}\int_{-L}^Ld{x}\left(\partial_t{n}+\partial_{{x}}{j}\right)^2$, which takes into account of the physical constraint, the conservation of particle number, as given in Eq.~(\ref{Eq:App1-nt}).

\begin{figure*}[ht]
\centering\includegraphics[width = 0.95\linewidth]{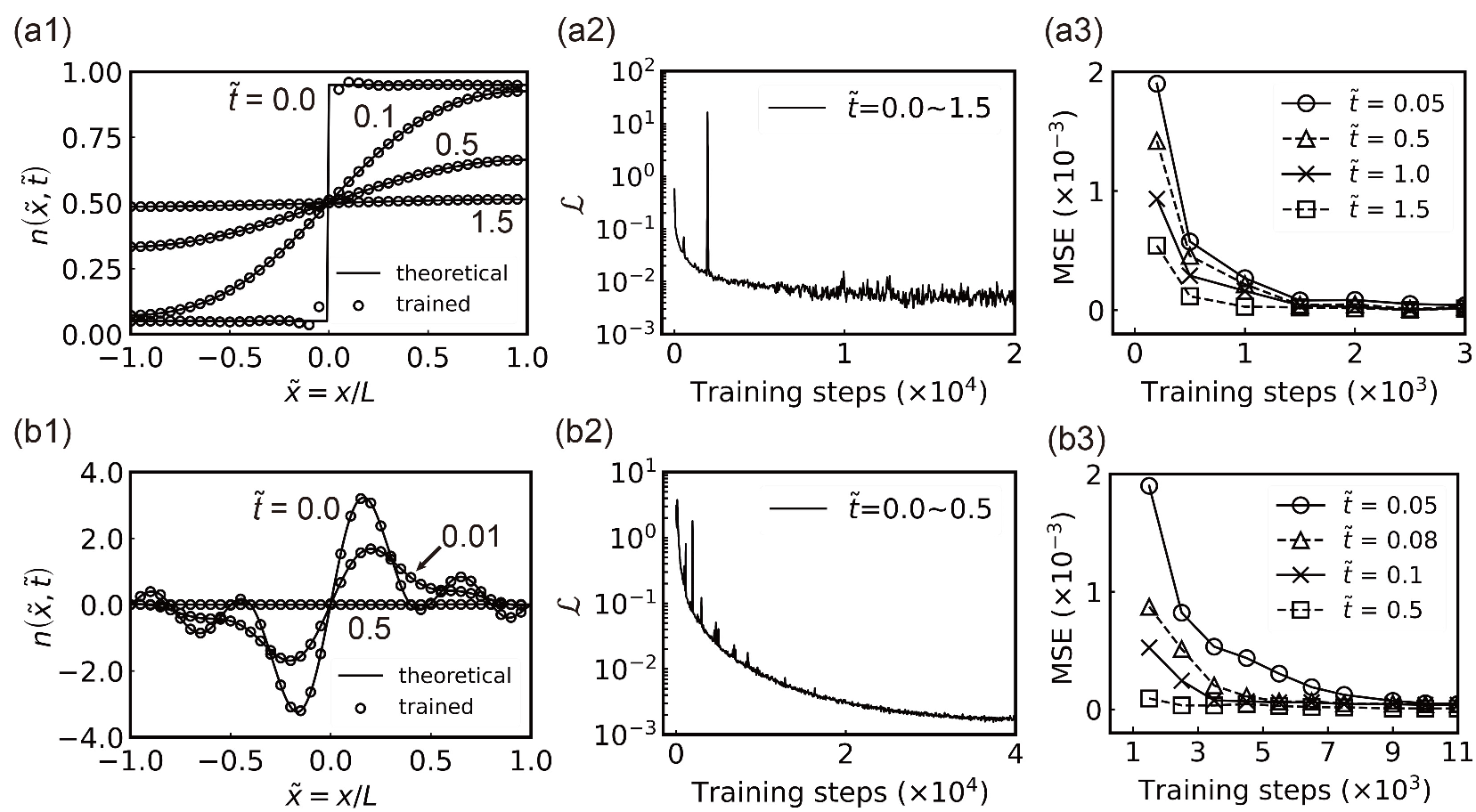}
\caption{Application 1 -- Solving 1D diffusion dynamics: comparison between the trained solution using DOMM (open circles, no time-splitting) and the exact solutions (solid lines) in Eqs.~(\ref{Eq:App1-Sol1}, \ref{Eq:App1-Sol2}) upon different initial and boundary conditions. (a1) Trained solution with an initial condition of a step function ${n}\left({x,0}\right)=0.5+0.45 \, \mathrm{sgn}\left({x}\right)$ and Neumann boundary condition at $x=\pm L$. (b1) Trained solution with an initial condition of sums of sinusoidal functions ${n}\left({x,0}\right)=\Sigma_{k=1}^{4} \sin(k\pi {x})$ and Dirichlet boundary condition at $x=\pm L$. The time $t$ is normalized by the unit $\tau_0=L^2/D$ and $\tilde{t}=t/\tau_0$. (a2) and (b2) show the convergence of the training loss $\mathcal{L}$ as a function of training steps for the solutions in (a1) and (b1), respectively. (a3) and (b3) display the convergence of Mean Squared Error (MSE) of particle density $n$ at different times as a function of training steps for the solutions to the two initial-boundary value problems. All the trained solutions are obtained using the same neural network structure with width $N_{\mathrm{w}}=50$, depth $N_{\mathrm{d}} = 6$. } \label{Fig:1dDiffusion} 
\end{figure*}

In this work, we use the Mean Squared Error (MSE) of the state variable $\alpha$, defined by $\frac{1}{N_{\mathrm{s}}}\sum_{i=1}^{N_{\mathrm{s}}}(\alpha_{\mathrm{trained}}-\alpha_{\mathrm{theoretical}})^2$, to measure the difference between the trained result ($\alpha_{\mathrm{trained}}$) and the theoretical result ($\alpha_{\mathrm{theoretical}}$, obtained analytically or numerically using traditional methods). Note that MSE is not used to inform the neural network or the training process when to stop, but is used only for accuracy checking (in comparison to results obtained from traditional ''exact" methods) and convergence analysis to show that the training process has enough accuracy and systematic convergence: as the training steps increase, the trained solution converges to the accurate solution systematically (almost monotonically). In Fig.~\ref{Fig:1dDiffusion}, we compare the trained solutions $n(x,t)$ using the DOMM and the exact solutions in Eqs.~(\ref{Eq:App1-Sol1}) and (\ref{Eq:App1-Sol2}) for the two cases at several different times, respectively. From this figure, we can see that the DOMM can solve the 1D diffusion dynamics with very high accuracy (the final MSE less than $10^{-5}$). Moreover, the errors and the loss function are both found to decay systematically with increasing training steps, implying training convergence. Here all the trained results are obtained using the same network structure with width $N_{\mathrm{w}}=50$, depth $N_{\mathrm{d}}=6$. All the weights ($w_{\mathrm{om}}, w_{\mathrm{con}}, w_{\mathrm{ic}}, w_{\mathrm{bc}}$) in the loss function are taken to be unity. 
In addition, in Fig.~\ref{Fig:1dDiffusion}-(a3) and Fig.~\ref{Fig:1dDiffusion}-(b3), we plot the decay of the MSEs with increasing training steps at several different times in the early stages of the training process for the two initial-boundary value problems, respectively. This not only demonstrates the convergence of the proposed DOMM method but also reveals an interesting frequency principle~\cite{Xu_2020}: A DNN tends to learn a target function from low to high frequencies during the training. The solution gets smoother (low frequencies in space) as the diffusion dynamics progresses and hence the solution at the later time is well-trained first. As shown in Fig.~\ref{Fig:1dDiffusion}-(a3) and Fig.~\ref{Fig:1dDiffusion}-(b3), although the trained solution near the earlier time has not yet converged to high accuracy, the computation accuracy at the later time is already very high. 

 \begin{figure*}[ht] 
\centering\includegraphics[width = 0.9\linewidth]{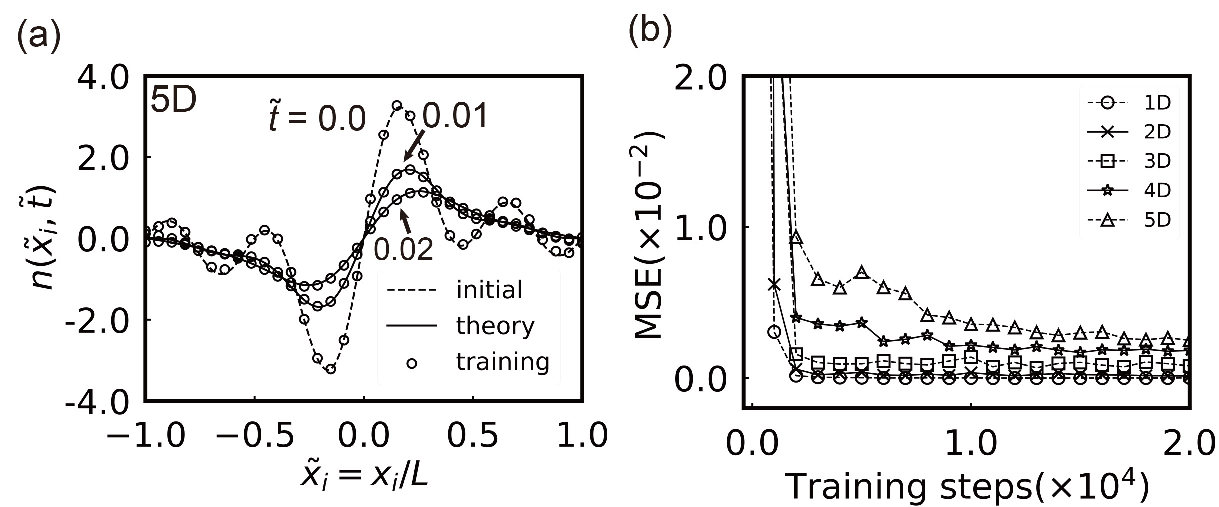}
\caption{Solving high-dimensional diffusion dynamics: comparison between the trained solution using DOMM (open circles, no time-splitting) and the analytical solutions (solid lines) in Eq.~(\ref{Eq:App1.5-Sol}) upon the initial condition in Eq.~(\ref{Eq:App1.5-n0}) and the Dirichlet boundary condition of $n(\{x_i=\pm L\},t)=0$ ($i=1,..,d$) for $\tilde{t}=t/\tau_0\in [0.0,0.02]$. (a) Slice plots of the trained solution and the analytical solution in the 5-dimensional system (by setting the value of $x_i$ non-zero only in one of the 5 dimensions). (b) The MSE convergence (or the accuracy) of the trained solutions at 1- to 5-dimensional systems is plotted as a function of the number of training steps. All the trained solutions are obtained using the same DNN structures and parameters (depth $5$, width $200$, learning rate $0.016$, dataset $10^4$). } \label{APP1.5-results} 
\end{figure*}

\subsection{Solving high-dimensional diffusion dynamics}\label{Sec:App1_2}

Based on the above successful implementation of DOMM to solve 1D diffusion dynamics, we now use DOMM to solve diffusion dynamics in high-dimensional spaces, by which we can better demonstrate the capability of DOMM in tackling high-dimensional problems. This type of high-dimensional problem is challenging for traditional methods (including those that minimize the OM integral with traditional optimization techniques) due to the curse of dimensionality, which makes it difficult to form a spatial grid with a sufficient number of grid points. However, in DOMM or other deep-learning methods for solving PDEs, we can efficiently utilize low-discrepancy sequences or quasi-Monte Carlo methods to handle such problems and obtain more accurate numerical solutions. 

Following the notations in Sec.~\ref{Sec:App1_1}, the particle number density $n(\mathbf{x},t)$ follows the conservation equation 
\begin{equation}\label{Eq:App1.5-nt}
\partial_t{n}+{\nabla\cdot \mathbf{j}}=0,
\end{equation}
where $\mathbf{x}=\{x_1,x_2,...,x_d\} \in [-L, L]^d$ in the general $d$-dimensional space with $d=2,3,4,5$, respectively, $\mathbf{j}(\mathbf{x},t)$ is the particle flux, and $\nabla$ is the $d$-dimensional gradient operator. Following the same procedure in Sec.~\ref{Sec:App1_1}, we find that the Onsager-Machlup integral given in Eq.~(\ref{Eq:App1-OM}) for 1-D diffusion is now given by
\begin{equation}\label{Eq:App1.5-OM}
% \mathcal{L}_{\mathrm{om}} = \int_0^{t}d{t} \left\{\mathcal{R}[j]-\mathcal{R}[j^*]\right\}= 
\mathcal{L}_{\mathrm{om}}=\int_0^{t}d{t}\int d{\mathbf{x}}\frac{\zeta}{2n}\left|{\mathbf{j}}-{\mathbf{j}^*}\right|^2,
\end{equation}
with the most probable flux $\mathbf{j}^{*}$ given by
\begin{equation}\label{Eq:App1.5-jstar}
\mathbf{j}^{*}=- D\nabla n.
\end{equation}
% with dot superscript denoting the time derivative and $D={k_{\mathrm{B}} T}/{\zeta}$ being the diffusion constant. 
Setting $\mathbf{j}=\mathbf{j}^{*}$ and substituting Eq.~(\ref{Eq:App1.5-jstar}) into the conservation equation~(\ref{Eq:App1.5-nt}) gives the $d$-dimensional diffusion equation
\begin{equation}\label{Eq:App1.5-1DEqn}
{\partial_t n}-D{\Delta n}=0,
\end{equation} 
with $\Delta$ being the $d$-dimensional Laplacian operator. Particularly, for the initial condition 
\begin{equation}\label{Eq:App1.5-n0}
{n}\left(x_i,0\right)=\Sigma_{k=1}^{4} \left[\sum_{i=1}^d \sin(k\pi x_i/L)\right],    
\end{equation}
and the Dirichlet boundary condition of $n(\{x_i=\pm L\},t)=0$ ($i=1,..,d$), the equation~(\ref{Eq:App1.5-1DEqn}) can be solved analytically to be  
\begin{equation}\label{Eq:App1.5-Sol}
n(\mathbf{x},t) = \sum_{k=1}^4 \exp\left({-k^2\pi^2t/\tau_0}\right)\left[\sum_{i=1}^d \sin(k\pi x_i/L)\right].
\end{equation}
with $\tau_0=L^2/D$. 

Similarly to the solution of 1-dimensional diffusion dynamics presented in Sec.~\ref{Sec:App1_1}, we use the DOMM to solve the above $d$-dimensional diffusion dynamics for $\tilde{t}=t/\tau_0\in [0.0,0.02]$. The same network parameters (depth $5$, width $200$, learning rate $0.016$, dataset $10^4$) are used during the solution training in the 2- to 5-dimensional diffusion problems. In Fig.~\ref{APP1.5-results}, we draw slice plots by sampling one dimension of the $d$-dimensional space (via setting the value of $x_i$ non-zero only in one of the dimensions). Fig.~\ref{APP1.5-results}(a) shows the high accuracy of trained solutions as shown by their very good agreement with analytical solution in Eq.~(\ref{Eq:App1.5-Sol}) in all high-dimensional systems (here we have only shown the $5$-dimensional results as an example). Fig.~\ref{APP1.5-results}(b) demonstrates the training convergence as indicated by decreasing MSE with training steps.  As the system dimension increases so does the complexity of solution training, the speeds of training convergence in all high-dimensional systems are still very fast with only a weak decrease with increasing system dimensions.  

In addition, in Table~\ref{Table:PINN_vs_DOMM}, we compare the performance of DOMM with PINN in solving 1- to 3-dimensional diffusion dynamics. By using the same DNN structure and parameters (depth $5$, width $60$, learning rate $0.016$, dataset $10^4$), we list the training steps needed for DOMM and PINN to reach the MSE value of $10^{-4}$, respectively. It can be seen that DOMM is much more efficient than PINN; only about half of the training steps in PINN are needed for DOMM to reach the same accuracy (MSE) of the trained solutions. Moreover, we also compared DOMM with PINN in terms of final accuracy and the average training time per 100 training steps. It can be observed that the proposed DOMM method consistently outperforms PINN for the solution of 1D to 3D diffusion dynamics in both final accuracy and training time (efficiency). In summary, all these results obtained for high-dimensional diffusion dynamics demonstrate the capability, efficiency, and convergence of DOMM as a new generic computational method to solve high-dimensional dynamic problems.
% More specifically, in terms of final accuracy, DOMM is more accurate than PINN by 1 digit, while in terms of training time, DOMM's training time is approximately $60\%$ to $80\%$ of PINN's.

\begin{table}[htbp]\color{black}
\centering
\renewcommand{\arraystretch}{2}
\setlength{\tabcolsep}{3 pt} 
\caption{Comparison of DOMM and PINN in solving 1D to 3D diffusion dynamics}\label{Table:PINN_vs_DOMM}
\vspace{5pt}
\begin{tabular}{|cc|c|ccc|}
\hline
\multicolumn{2}{|c|}{\multirow{2}{*}{\bf{Item}}}                                            & \multirow{2}{*}{\bf{Method}} & \multicolumn{3}{c|}{\bf{Dimension}}                                                       \\ \cline{4-6} 
\multicolumn{2}{|c|}{}                                                                 &                         & \multicolumn{1}{c|}{\bf{1D}}           & \multicolumn{1}{c|}{\bf{2D}}           & {\bf{3D}}          \\ \hline
\multicolumn{1}{|c|}{\multirow{4}{*}{\bf{Efficiency}}} & \multirow{2}{*}{\makecell[c]{\bf{Convergence speed} \\ (training steps required \\ when MSE$\leq 10^{-2}$)}} & DOMM                    & \multicolumn{1}{c|}{\bf{\underline{1,250 steps}}}  & \multicolumn{1}{c|}{\bf{\underline{5,200 steps}}}  & \bf{\underline{23,100 steps}} \\ \cline{3-6} 
\multicolumn{1}{|c|}{}                            &                                    & PINN                    & \multicolumn{1}{c|}{2,350 steps}  & \multicolumn{1}{c|}{9,350 steps}  & 39,550 steps \\ \cline{2-6} 
\multicolumn{1}{|c|}{}                            & \multirow{2}{*}{\makecell[c]{\bf{Training time} \\ (per 100 steps)}}     & DOMM                    & \multicolumn{1}{c|}{\bf{\underline{1.44 seconds}}} & \multicolumn{1}{c|}{\bf{\underline{2.19 seconds}}} & \bf{\underline{3.54 seconds}} \\ \cline{3-6} 
\multicolumn{1}{|c|}{}                            &                                    & PINN                    & \multicolumn{1}{c|}{2.50 seconds} & \multicolumn{1}{c|}{3.16 seconds} & 4.17 seconds \\ \hline
\multicolumn{1}{|c|}{\multirow{2}{*}{\bf{Accuracy}}}   & \multirow{2}{*}{\makecell[c]{\bf{Final accuracy} \\ (MSE after $10^{5}$ steps)}}    & DOMM                    & \multicolumn{1}{c|}{\underline{$\mathbf{10^{-7}}$}}          & \multicolumn{1}{c|}{$10^{-5}$}           & {\underline{$\mathbf{10^{-5}}$}}            \\ \cline{3-6} 
\multicolumn{1}{|c|}{}                            &                                    & PINN                    & \multicolumn{1}{c|}{$10^{-6}$}           & \multicolumn{1}{c|}{$10^{-5}$}           & {$10^{-4}$}           \\ \hline
\end{tabular}
\end{table}

\subsection{Some auxiliary strategies to improve the accuracy and convergence}

We now introduce the auxiliary training strategies used to improve the accuracy and accelerate the convergence of the DOMM in solving diffusion dynamics in dilute solutions and the following applications to two-phase dynamics.

(i) \emph{Optimization of DNN structure and activation function}. There are many different types of DNN structures, for example, Fully-connected Neural Networks (FNNs), Residual Neural Networks (ResNet)~\cite{he2016deep}, and Convolutional Neural Networks (CNNs)~\cite{krizhevsky2012imagenet}. Moreover, many different activation functions have also been employed such as $\mathrm{ReLU}$, $\mathrm{ReLU}^3$, $\mathrm{Tanh}$~\cite{Weinan2018}. During the training process in our dynamic problems such as diffusion and two-phase dynamics, we observe that the combination of ResNet and $\mathrm{Tanh}^3$ consistently outperforms other combinations of network structures and activation functions. This combination also successfully avoids issues like gradient explosion or vanishing. Therefore, in this work, we employ the combination of ResNet and $\mathrm{Tanh}^3$ in all numerical cases to showcase the results. Further exploration of the performance of different network structures and activation functions will be left for future work. The major structure components of the DNN to be tuned and optimized are the network depth $N_{\mathrm{d}}$, the network width $N_\mathrm{w}$, the number of sampling data (observation) points $N_{\mathrm{s}}$, the activation functions defined in Eq.~(\ref{Eq:DOMM-MLNet}), the learning rate $\eta_{\mathrm{r}}$ defined in Eq.~(\ref{Eq:DOMM-SGD}), the optimization method, and the number of training steps.
% epochs (an epoch represents one pass through the entire training dataset). 
Moreover, we observe that increasing the number of training steps improves the performance of the DOMM in learning complex forms of solutions, as in other DNN methods for solving PDEs. 

\begin{figure}[htbp]
\centering
\includegraphics[width = 0.65 \linewidth]{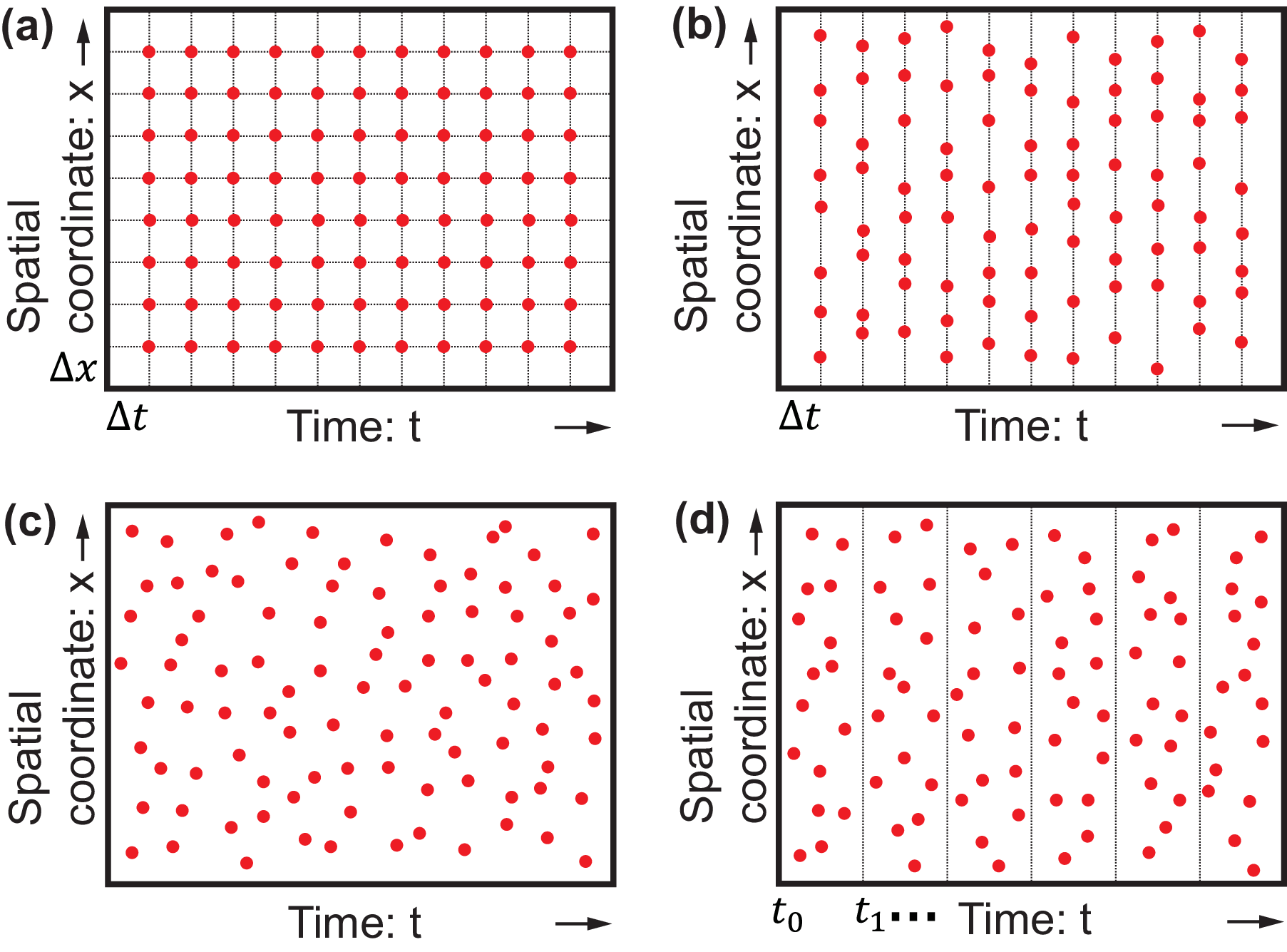}
\caption{Comparisons of different sampling strategies used in the DNN methods for solving PDEs. (a) Discrete sampling on the spatiotemporal meshes as in the traditional finite difference method (FDM) and finite element method (FEM)~\cite{ChunLiu2022,Reina2021VONN}. (b) Random (uniform) sampling in spatial coordinates on a discrete mesh of small time steps: mesh-free only in space. (c) Random (uniform) sampling in both spatial and temporal coordinates: mesh-free in both space and time. (d) Random (uniform) sampling in both spatial and temporal coordinates. Multi-step training is carried out by splitting the whole time period into several time intervals.}
\label{Fig:SamplingMethods}
\end{figure}

(ii) \emph{Sampling strategy: Time splitting}. There are many different sampling methods in DNN methods for solving PDEs as shown in Fig.~\ref{Fig:SamplingMethods}, for example, (a) the discrete sampling on the spatiotemporal meshes as in the traditional finite difference
method (FDM) and finite element method (FEM)~\cite{ChunLiu2022,Reina2021VONN}, (b) the random (uniform) sampling in spatial coordinates on a discrete mesh of small time steps, (c) the random (uniform) sampling in both spatial and temporal coordinates over the whole spatial domain and time period, (d) the random (uniform) sampling in both spatial and temporal coordinates, where multi-step training is carried out by splitting the whole time period into several time intervals.  

When we use DOMM to solve the 1D diffusion and two-phase dynamics, we observe that in comparison to the sampling on regular meshes of spatial and/or temporal coordinates as in (a) and (b), the mesh-free random sampling strategy (c) exhibits many advantages: the training is relatively insensitive to the order of derivatives in the OM functionals and the training efficiency shows a much slower increase with the increasing number $N_{\mathrm{s}}$ of spatiotemporal coordinate (input) points. Furthermore, we observe that the most efficient sampling method is the combination of the mesh-free random sampling strategy with the time-splitting strategy (d). In this strategy, instead of performing training over the entire target long time period (say, from $t_0$ to $t_{\mathrm{f}}$), the time interval is divided into several smaller intervals (\emph{e.g.}, $t_0,t_1=t_0+\Delta t,\ldots,t_{N}=t_{\mathrm{f}}$) with equal time step $\Delta t$. Training is then performed step by step, such as from $t_0$ to $t_1$, then from $t_1$ to $t_2$, and so on until $t_{N-1}$ to $t_{N}$. Such a sampling strategy of time-splitting reduces the cost of computational resources significantly and enables training of high accuracy through a much smaller-scale network (and less computational resource) over a given time period.

We observe that during the training, the solutions in the latter time are well-trained first, as shown in Fig.~\ref{Fig:1dDiffusion}-(a3) and Fig.~\ref{Fig:1dDiffusion}-(b3), that is, for the training step from $t_i$ to $t_{i+1}$, the solution at time $t_{i+1}$ is well trained much more quickly than that at time $t_i$. This observation is consistent with the frequency principle proposed by Xu \emph{et al}~\cite{Xu_2020}: A DNN tends to learn a target function from low to high frequencies during the training. In the soft matter dynamic processes considered in this work such as the diffusion dynamics, the solutions get smoother (low frequencies or smaller wavenumber) during temporal evolution and hence the solution at a later time will be well-trained first. Furthermore, we would like to remark that this observation tells us that the training of DNNs for the solution of a time-dependent PDE is significantly different from traditional numerical methods where the historical solution is necessary for time marching. Here in DOMM, there is no time-marching step in the training process, and one can obtain an accurate solution at a later time even though the solutions at earlier times are still far away from accurate solutions. However, herein, we only consider dynamic processes (such as diffusion dynamics) where the solutions get smoother with time evolution. It is still not clear and interesting to see in the future whether the frequency principle also works for dynamic processes where the solution gets sharper with increasing time, for example, Burgers equation with shock formation.

In addition, in Table~\ref{Table:1dDiffusion}, we present detailed comparisons between the training algorithms using sampling strategies (c, no splitting) and (d, with time splitting), as illustrated in Fig.~\ref{Fig:SamplingMethods}, in solving 1D diffusion dynamics, highlighting the time-splitting sampling strategy's faster convergence and higher accuracy. Algorithm I represents the training without time splitting, where the training time spans the entire target time period from $t/\tau_0 = 0.0$ to $1.5$. On the other hand, Algorithm II utilizes time splitting, dividing the training into three time intervals: $0.0-0.5$, $0.5-1.0$, and $1.0-1.5$, with a splitting time step $\Delta t/\tau_0= 0.5$. Moreover, during each training interval, Algorithm II employs much fewer (one-third) sampling data points than those used in Algorithm I. In the table, the ``step" column indicates the number of training steps required for reaching an MSE magnitude below $5.0 \times 10^{-5}$ for the corresponding time interval and algorithm. Additionally, MSE$_{\mathrm{f}}$ represents the final MSE value after $20,000$ training steps. We observe that the use of time splitting in Algorithm II endows a faster decrease in MSE while utilizing only one-third of the data compared to Algorithm I. For example, for the time interval $t/\tau_0=0.0-0.5$, Algorithm I requires approximately $2,700$ training steps to achieve an MSE below $5.0 \times 10^{-5}$, whereas Algorithm II with time splitting achieves the same level of accuracy in approximately $2,100$ steps. Similar conclusions hold for the time intervals $0.5-1.0$ and $1.0-1.5$. It is also important to note that although the MSE below $5.0 \times 10^{-5}$ seems to indicate high accuracy, further training is still necessary. The subsequent training, despite potentially yielding only marginal improvements in the MSE, plays a significant role due to the frequency nature of neural network training. Moreover, it can also be observed that in Algorithm I, the numbers of training steps required to reach the same accuracy level corresponding to $t/\tau_0=0.5-1.0$ and $t/\tau_0=1.0-1.5$ are smaller than those required for $t/\tau_0=0.0-0.5$. This once again highlights this phenomenon in the training process. Furthermore, time splitting not only accelerates training to some extent but also improves the overall precision of the training as observed from MSE$_{\mathrm{f}}$. This signifies the effectiveness of the time-splitting strategy. 
\begin{table}[htbp]
\centering
% \title{Solving 1D diffusion dynamics with time-splitting strategy}
  \renewcommand{\arraystretch}{1.5} % 调整行间距为原来的 1.5 倍
  \setlength{\tabcolsep}{10 pt} % 调整列间距为 12pt
  \caption{Solving 1D diffusion dynamics}\label{Table:1dDiffusion}
  \vspace{8pt}
  \begin{tabular}{|l|ll|ll|}
    \hline
    & \multicolumn{2}{c|}{\bf{Algorithm I}} & \multicolumn{2}{c|}{\bf{Algorithm II}} \\ \hline
    \multicolumn{1}{|c|}{$t/\tau_0$} & \multicolumn{1}{l|}{step} & \multicolumn{1}{c|}{MSE$_{\mathrm{f}}$} & \multicolumn{1}{l|}{step} & \multicolumn{1}{c|}{MSE$_{\mathrm{f}}$} \\ \hline
    0.0-0.5 & \multicolumn{1}{l|}{2700} & 9.4048 & \multicolumn{1}{l|}{2100} & 5.7967 \\ \hline
    0.5-1.0 & \multicolumn{1}{l|}{700}  & 1.0893 & \multicolumn{1}{l|}{400}  & 0.070390 \\ \hline
    1.0-1.5 & \multicolumn{1}{l|}{700}  & 0.79119 & \multicolumn{1}{l|}{400}  & 0.012141 \\ \hline
  \end{tabular}
  % \caption{Solving 1D diffusion dynamics: Faster convergence and higher accuracy granted by the time-splitting sampling strategy. Algorithm I (no time splitting): one-step training spans the entire target time period from $t/\tau_0 = 0.0$ to $1.5$. Algorithm II (employing the time-splitting strategy): three-step training by dividing the training into three-time intervals ($0.0-0.5$, $0.5-1.0$, and $1.0-1.5$). The ``step" column indicates the number of training steps needed to achieve the accuracy of the magnitude of MSE below $5.0 \times 10^{-5}$ for each algorithm and time interval. The MSE$_{\mathrm{f}}$ column lists the final MSE value (with units $10^{-6}$) after the training of $20,000$ steps for each algorithm and time interval.} 
\end{table}

(iii) \emph{Mini-batch training strategy.} Mini-batching is a technique that has been widely used in deep learning to improve performance~\cite{Yuan2018,zhao2021PINN}. In the optimization process, it suggests not calculating the exact direction of the gradient using the entire sampled full dataset but dividing the dataset into many subsets, called mini-batches. Mini-batching has been shown to help avoid less desirable local minimum better than full-batch gradient descent~\cite{Yuan2018}. In this work, we adopt the same mini-batch training strategy as in the deep Ritz method (DRM)~\cite{Weinan2018}, where the mini-batching is claimed to be able to prevent overfitting issues that may arise as the training progresses with some specified quadrature points. Here we observe that the use of mini-batches introduces more fluctuations in the optimization of the loss function during the training process and promotes faster convergence of DNNs. 

(iv) \emph{Adjusting the loss-function weights.} Adjusting the weights in the loss functions is subtle in most DNNs methods for solving PDEs. A rule of thumb is that we choose the weight of a loss-function term to keep its value to be the same order of magnitude as other terms such that all the loss-function terms can be optimized almost simultaneously. In our computations using DOMM (with loss functions taking the form of Eq.~(\ref{Eq:DOMM-Loss})), we find that the weights of loss-function terms associated with initial conditions and physics/geometric constraints are usually needed to take larger values than other weights. 

(v) \emph{Imposing boundary conditions.} We remark that there are many different approaches to imposing boundary conditions or constraints when using DNNs to solve problems related to differential operators. Here we employed the penalty method, one of the most general and commonly used soft-constraint methods. There exist many other methods that apply to specific constraints and can be more efficient, for example, the hard-constraint methods that impose boundary conditions through certain DNN constructions~\cite{SUKUMAR2022114333}. Using such methods, the Dirichlet boundary condition can be imposed by modifying the training output layer through the multiplication of a boundary distance function. Similarly, periodic boundary conditions can be enforced by modifying the input layer properly. However, these hard-constraint construction methods become very complicated and inefficient when dealing with problems with complex computational domains and boundary conditions.

\section{Application 2: Solving two-phase dynamics}\label{Sec:App2} 

\subsection{Two-phase dynamics: Cahn-Hilliard equation} 

Firstly, the conservation equation of $\phi$ reads
\begin{equation}\label{Eq:App2-phit}
\partial_t{\phi}+{\nabla \cdot \mathbf{J}}=0,
\end{equation}
where $\mathbf{J}(\mathbf{x},t)$ is the diffusion flux. The free energy of the system is given by 
\begin{equation}\label{Eq:App2-F}
\mathcal{F} [\phi(\mathbf{x},t)]=\int d\mathbf{x} \left[\frac{a}{4}\left(\phi^2-1\right)^2+\frac{K}{2} \left(\nabla \phi \right)^2\right],
\end{equation}
where $a$ is a positive constant in the double-well potential and $K$ is the positive interfacial stiffness parameter. The interfacial thickness and interfacial tension are given by $\epsilon=\sqrt{K/a}$ and $\gamma=2\sqrt{2} a\epsilon /3$, respectively. The dissipation function $\Phi$ takes the following quadratic form
\begin{equation}\label{Eq:App2-Phi}
\Phi[\mathbf{J}(\mathbf{x},t)]=\int d\mathbf{x} \frac{\mathbf{J}^{2}}{2M},
\end{equation}
with $M$ being the constant mobility. Minimizing the Rayleighian $\mathcal{R}[\mathbf{J}(\mathbf{x},t)]=\dot{\mathcal{F}}+\Phi$ with respect to the flux $\mathbf{J}$ yields the most probable flux 
\begin{equation}\label{Eq:App2-Jstar}
\mathbf{J}^{*}=-M\nabla \hat{\mu},
\end{equation} 
with $\hat{\mu}$ being the generalized chemical potential given by
\begin{equation}\label{Eq:App2-mu}
\hat{\mu} \equiv a\phi(\phi^2-1)-K\nabla^2 \phi.
\end{equation}
Substituting Eqs.~(\ref{Eq:App2-Jstar}) and (\ref{Eq:App2-mu}) into the conservation equation (\ref{Eq:App2-phit}) gives the standard Cahn-Hilliard equation
\begin{equation}\label{Eq:App2-CH}
\partial_t{\phi}-M\nabla^2 [a\phi (\phi^2-1)-K\nabla^2 \phi]=0,
\end{equation}
and the Onsager-Machlup integral is given by
\begin{equation}\label{Eq:App2-OM}
% \mathcal{L}_{\mathrm{om}} = \int_0^{t}d{t} \left\{\mathcal{R}[j]-\mathcal{R}[j^*]\right\}= 
\mathcal{L}_{\mathrm{om}}= \int d{t}\int d\mathbf{x}\frac{1}{2M}\left(\mathbf{J}-{\mathbf{J}^*}\right)^2.
\end{equation}

After showing the success of the DOMM in solving the simple 1D diffusion dynamics, we now turn to the solution of the more complicated two-phase dynamics with a conserved phase parameter $\phi$, which can be described by the Cahn-Hilliard equation that includes high-(fourth-) order derivatives~\cite{Doi2013,Qian2006}. 

\subsection{Solving two-phase dynamics in 1D: Cosine initial condition} 

We first consider the solution of two-phase dynamics in 1D~\cite{zhao2021PINN}. In this case, the state variable is $\phi(x,t)$, the phase field parameter at position $x$ and time $t$ and the diffusion flux is one-dimensional, $\mathbf{J}=J\mathbf{\hat{x}}$.
To be specific, we use DOMM to solve the 1D two-phase dynamics for $-L\leqslant x \leqslant L$ and $t \geqslant 0 $, where we take the Cosine initial condition, $\phi_0(x)=-\cos (2 \pi x/L)$, and the periodic boundary condition, $\phi(-L,t) =\phi(L, t)$ and $\partial_x{\phi}(-L,t) = \partial_x{\phi}(L,t)$. 

\begin{figure}[!ht]
\centering
\includegraphics[width = 0.8 \linewidth]{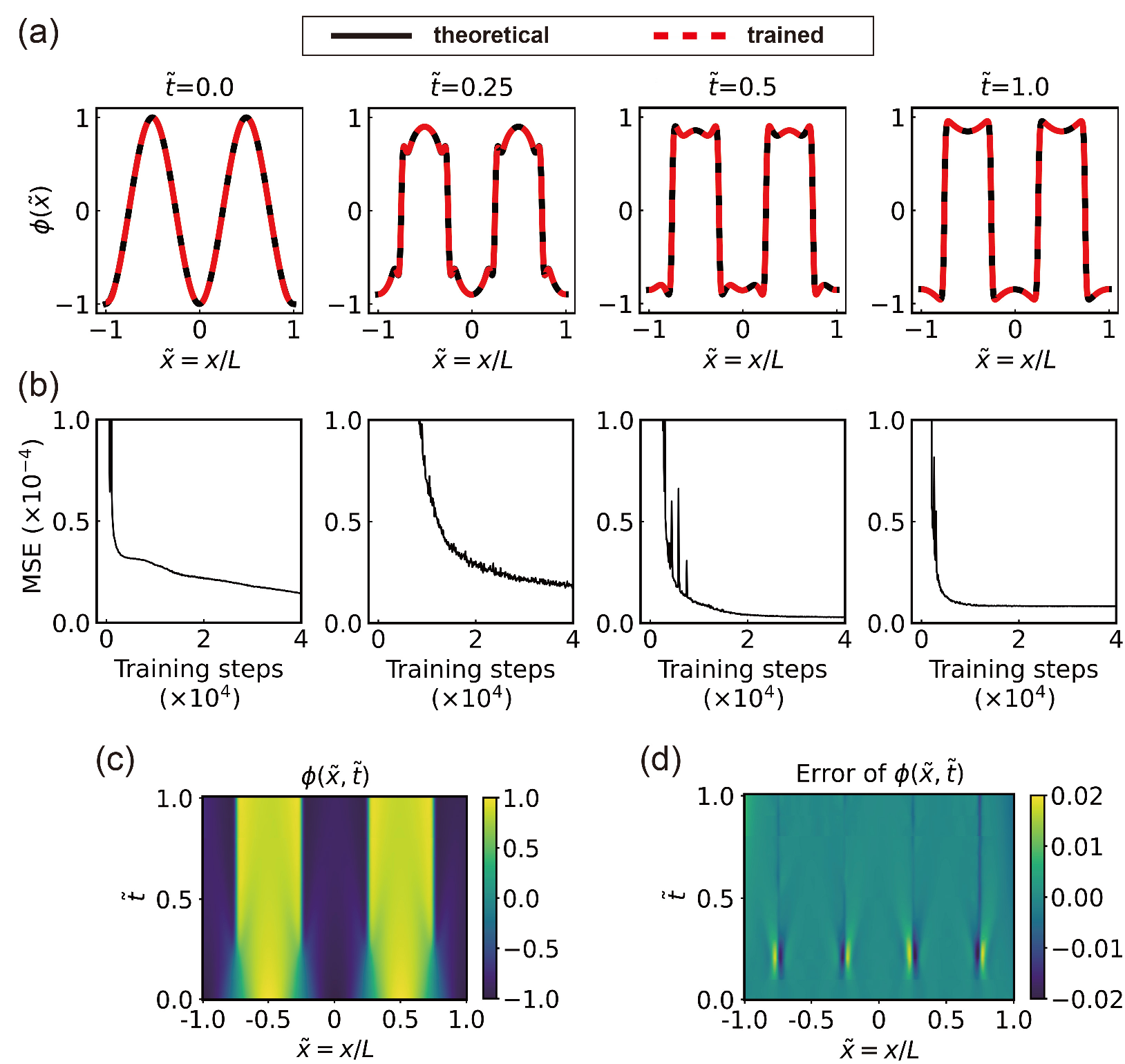}
\caption{Application 2 -- Solving 1D two-phase dynamics. (a) Comparison between the trained solution using DOMM (dashed lines) and the numerical solution using the traditional FDM (solid lines) upon the initial condition $\phi(x;t=0)=-\cos (2 \pi x/L)$ and periodic boundary condition at $x=\pm L$. The time $t$ is normalized by the unit $0.01\tau$ with $\tau\equiv L^2/Ma$ and the interface thickness is taken to be $\epsilon = 10^{-2}L$. 
(b) Convergence of the training error (Mean Square Error, MSE) of $\phi$ (at the corresponding time in (a)) as a function of the training steps. (c) Trained solution of $\phi$ as a function of position $\tilde{x}=x/L$ and time $\tilde{t}=t/\tau$. (d) The pointwise error maps of the trained solution of $\phi$ is calculated using $\phi_{\mathrm{DOMM}}(x,t) - \phi_{\mathrm{FDM}}(x,t)$ as a function of position and time with of $\phi_{\mathrm{DOMM}}$ being the trained solution by DOMM and $\phi _{\mathrm{FDM}}$ being the numerical solution by the traditional FDM. All the trained results are obtained using the same neural network structure with width $N_{\mathrm{w}}=128$, depth $N_{\mathrm{d}}=3$.}
\label{Fig:1dCH}
\end{figure}

To use the DOMM to solve the 1D two-phase dynamics, the input parameters of the neural network are still the position $x$ and time $t$; the output parameters are chosen to be the phase-field parameter $\phi$ and the diffusion flux $J$. The loss function minimized to train the neural network still takes the general form of Eq.~(\ref{Eq:DOMM-Loss}). Here the first loss function term $\mathcal{L}_{\mathrm{om}}$ is the Onsager-Machlup integral given in Eq.~(\ref{Eq:App2-OM}) with $\mathbf{J}=J\mathbf{\hat{x}}$, $\mathbf{J}^*=J^*\mathbf{\hat{x}}$ given in Eq.~(\ref{Eq:App2-Jstar}), and $J^*=-M\partial_x \hat{\mu}$. The second loss function term is associated with the periodic boundary condition: 
\begin{equation}\label{Eq:App2-1DCH-LossBC}
\begin{aligned}
\mathcal{L}_{\mathrm{bc}}=&\int_0^{t}d{t'} \left\{\left[\phi(-L,t') - \phi(L, t')\right]^2\right.\left.+L^2\left[\partial_x {\phi}(-L,t') - \partial_x {\phi}(L,t')\right]^2\right\}.
\end{aligned}
\end{equation}
% The third loss function term imposing the initial condition is given by $\mathcal{L}_{\mathrm{ic}}=\int_{-L}^Ld{x}\left[\phi\left(x, 0\right)-\phi_0\right]^2$. The last loss function term is given by $\mathcal{L}_{\mathrm{con}}=\int_0^{t}d{t'}\int_{-L}^Ld{x} \left(\partial_t{\phi}+\partial_{{x}}J\right)^2$, which takes into account of the physical constraint of $\phi$ conservation, as given in Eq.~(\ref{Eq:App2-phit}). 
The third loss function term imposing the initial condition is $\mathcal{L}_{\mathrm{ic}}=\int_{-L}^Ld{x}\left[\phi\left(x, 0\right)-\phi_0\right]^2$. The last loss function term is given by $\mathcal{L}_{\mathrm{con}}=\int_0^{t}d{t'}\int_{-L}^Ld{x} \left(\partial_t{\phi}+\partial_{{x}}J\right)^2$, which takes into account of the physical constraint of $\phi$ conservation, as given in Eq.~(\ref{Eq:App2-phit}). Note that, compared to PINN directly using the fourth-order Cahn-Hilliard equation, the proposed DOMM method is based on the OM integral where only the third-order derivative appears. This, then, results in reduced use of automatic differentiation and helps to save computational costs.

Firstly, we note that for the artificially-designed 1D two-phase dynamics, the solution starts from a Cosine initial condition and gets sharper during the time evolution (high frequencies or large wavenumber), which is in contrast to the diffusion dynamics studied in Sec.~\ref{Sec:App1}. In this case, we observe that during the training, the solutions in the earlier time (smoother with lower frequencies) are well-trained first, that is, for the training step from $t_i$ to $t_{i+1}$, the solution at time $t_{i}$ is well trained much more quickly than that at time $t_{i+1}$. This observation is again consistent with the frequency principle proposed by Xu \emph{et al}~\cite{Xu_2020}: A DNN tends to learn a target function from low to high frequencies during the training. We would expect further that the frequency principle should also work in other more physical dynamic processes such as the shock formation described by Burgers equation where the solution gets sharper with increasing time.

Secondly, in Fig.~\ref{Fig:1dCH}, we compare the trained solutions $\phi(x,t)$ using the DOMM and the numerical solutions using the traditional FDM at several different times. From this figure, we can see that the DOMM can solve the 1D two-phase dynamics with high accuracy and convergence. The trained solutions are very close to the numerical solutions with final MSE less than $10^{-5}$, implying very high accuracy. Moreover, the errors and the loss function are both found to decay systematically with increasing training steps, ensuring training convergence. All the trained results are obtained using the same network structure with width $N_{\mathrm{w}}=128$, depth $N_{\mathrm{d}}=3$. We choose the weights in the loss function to be $w_{\mathrm{ic}} = 100.0$ and $w_{\mathrm{om}} = w_{\mathrm{con}} = w_{\mathrm{bc}} = 1.0$. Moreover, we have also used the sampling strategy of time splitting with time interval $\Delta t = 0.1$ to accelerate the training of the 1D two-phase dynamic process. In contrast, we solve the above problems using PINN with exactly the same hyper-parameters and sampling settings. The comparisons between DOMM and PINN are presented in Table~\ref{Table:PINN_vs_DOMM_CH}. It is notable that, similar to the results of diffusion dynamics, DOMM not only reduces the training steps required to achieve the same MSE but also consumes less training time per 100 steps compared to PINN, while yielding a higher final accuracy.

\begin{table}[htbp]\color{black}
\centering
\renewcommand{\arraystretch}{2}
\setlength{\tabcolsep}{4 pt} 
\caption{Comparison of DOMM and PINN in solving two-phase dynamics in 1D}\label{Table:PINN_vs_DOMM_CH}
\vspace{5pt}
\begin{tabular}{|cc|cc|}
\hline
\multicolumn{2}{|c|}{\multirow{2}{*}{\textbf{Item}}}                                    & \multicolumn{2}{c|}{\textbf{Method}}                       \\ \cline{3-4} 
\multicolumn{2}{|c|}{}                                                                  & \multicolumn{1}{c|}{\textbf{DOMM}}         & \textbf{PINN} \\ 
\hline
\multicolumn{1}{|c|}{\multirow{2}{*}{\textbf{Efficiency}}} & \makecell[c]{\bf{Convergence speed} \\ (training steps required \\ when MSE$\leq 10^{-2}$)} & \multicolumn{1}{c|}{\underline{\textbf{5,100 steps}}}  & 10,300 steps   \\ \cline{2-4} 
\multicolumn{1}{|c|}{}                                     & \makecell[c]{\bf{Training time} \\ (per 100 steps)}     & \multicolumn{1}{c|}{\underline{\textbf{3.17 seconds}}} & 4.56 seconds  \\ \hline
\multicolumn{1}{|c|}{\textbf{Accuracy}}                    & \makecell[c]{\bf{Final accuracy} \\ (MSE after $10^{5}$ steps)}   & \multicolumn{1}{c|}{\underline{{$\mathbf{10^{-5}}$}}}           & {$10^{-3}$}            \\ \hline
\end{tabular}
\end{table}

\subsection{Solving two-phase dynamics in 2D: Droplet coalescence} 

Now we turn to the solution of two-phase dynamics in 2D. In this case, the state variable is $\phi(x,y,t)$, the phase field parameter at position $\mathbf{x}=(x,y)$ and time $t$. To be specific, we consider the coalescence of two droplets (neglecting the flows)~\cite{zhao2021PINN} for $-L\leqslant x,\,y \leqslant L$ and $t \geqslant 0$, where we take the periodic boundary conditions and the following initial condition
\begin{equation}\label{Eq:App2-2DCH-ic}
\phi_0(x,y)=\max \left(\tanh \left(\frac{R_0-r_1}{2\epsilon}\right), \tanh \left(\frac{R_0-r_2}{2\epsilon}\right) \right),
\end{equation}
where $R_0 = 0.4 L$ is the droplet radius; the centers of the two droplets are located at $+0.7R_0 \mathbf{\hat{x}}$ and $-0.7R_0 \mathbf{\hat{x}}$, respectively; $r_1 = \sqrt{(x-0.7R_0)^2 + y^2}$ and $r_2 = \sqrt{(x+0.7R_0)^2 + y^2}$ are the distances of any space position $\mathbf{x}=(x,y)$ from the centers of the two droplets, respectively.

\begin{figure}[!ht]%bp
\centering
\includegraphics[width = 0.9\linewidth]{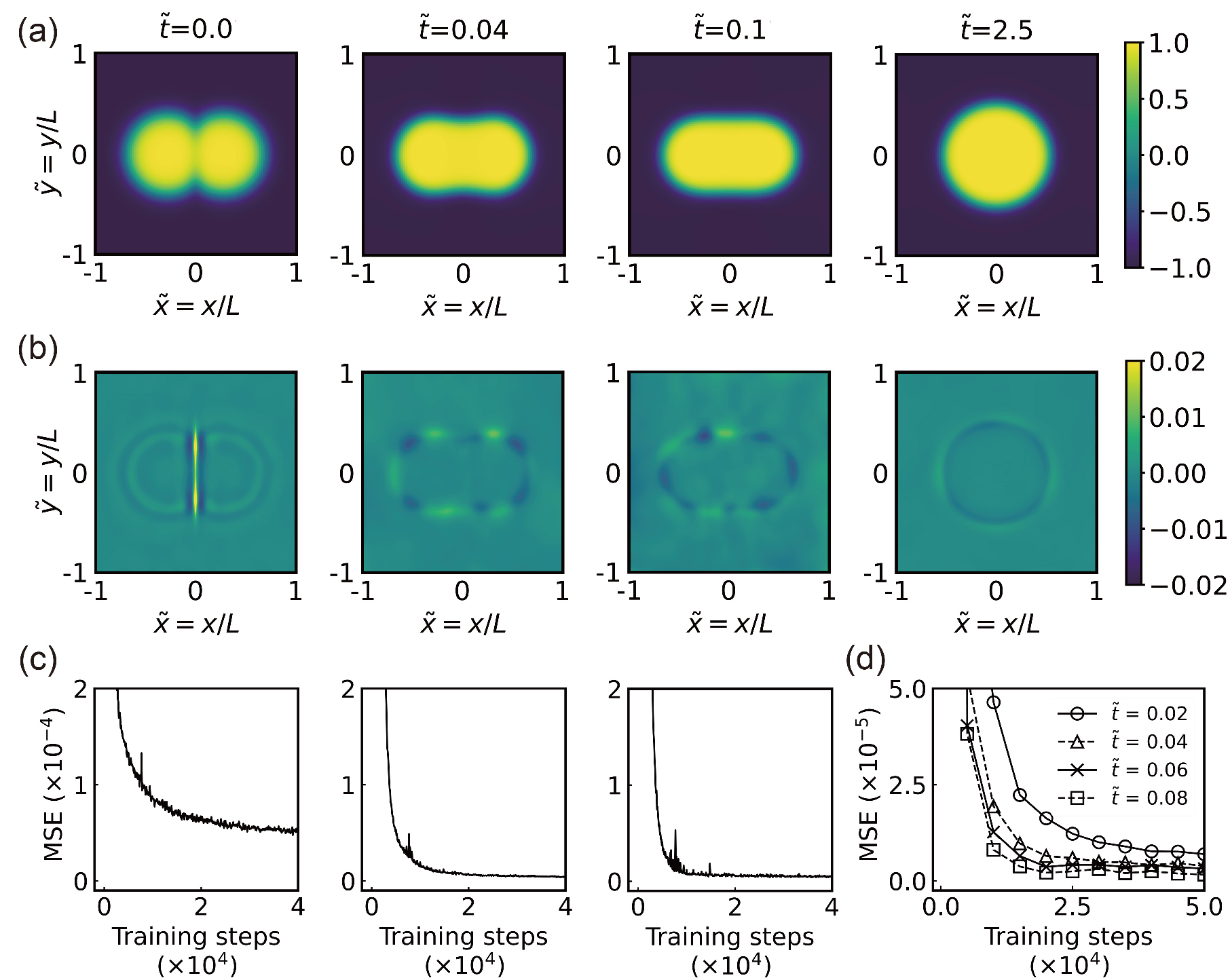}
\caption{Application 2 -- Solving 2D two-phase dynamics: coalescence of two droplets upon periodic boundary condition at $x,\, y=\pm L$. (a) Trained dynamical process of the two-droplets coalescence using DOMM (with time splitting). The trained result shows the two smaller droplets merge with each other gradually and form a larger circular droplet finally. The time $t$ is normalized by the unit $\tau\equiv L^2/Ma$ and the interface thickness is taken to be $\epsilon = 0.05 L$. (b) The pointwise error maps of the trained solution of $\phi$ (corresponding to the time in (a)) are calculated using $\left(\phi_{\mathrm{DOMM}}(\mathbf{x},t) - \phi_{\mathrm{FDM}}(\mathbf{x},t)\right)$ as a function of position and time with $\phi_{\mathrm{DOMM}}$ being the trained solution by DOMM and $\phi _{\mathrm{FDM}}$ being the numerical solution by traditional FDM. 
% These results further demonstrate the capability of DOMM in computing two-phase dynamics with high accuracy. 
(c) Convergence of MSE as a function of training steps for the solutions at $\tilde{t}=0.0, 0.04$, and $0.1$. (d) Convergence of MSE as a function of training steps for the solutions within the same training time interval $(\tilde{t} \in [0,0.1])$. The interface of coalescing droplets becomes smoother as the dynamic process progresses. During the training process, states with smoother interfaces tend to converge to better accuracy first, consistent with the empirical frequency principle~\cite{Xu_2020}. All trained results are obtained using the same neural network structure with width $N_{\mathrm{w}}=128$, depth $N_{\mathrm{d}}=3$.}
\label{Fig:2dCH}
\end{figure}

To use the DOMM to solve the 2D two-phase dynamics, the input parameters of the neural network are the position vector $\mathbf{x}=(x,y)$ and time $t$; the output parameters are chosen to be the phase-field parameter $\phi$ and the diffusion flux $\mathbf{J}=(J_x,J_y)$. The loss function minimized to train the neural network still takes the general form of Eq.~(\ref{Eq:DOMM-Loss}). Here the first loss function term $\mathcal{L}_{\mathrm{om}}$ is the Onsager-Machlup integral in Eq.~(\ref{Eq:App2-OM}) with $\mathbf{J}^*$ given in Eq.~(\ref{Eq:App2-Jstar}). The second loss function term is associated with the periodic boundary conditions:
\begin{equation}\label{Eq:App2-2DCH-LossBC}
\begin{aligned}
\mathcal{L}_{\mathrm{bc}}&=\int_0^{t}d{t'} \left\{\int_{-L}^{L}d{x} \left[\phi(x, -L, t') - \phi(x, L, t')\right]^2 \right.\\
&+\int_{-L}^{L}d{x} L^2 \left[\partial_x {\phi}(x, -L, t') - \partial_x {\phi}(x, L, t')\right]^2\\
&+\int_{-L}^{L}d{y} \left[\phi(-L, y, t') - \phi(L, y, t')\right]^2\\
&\left.+\int_{-L}^{L}d{y} L^2\left[\partial_y {\phi}(-L, y, t') - \partial_y {\phi}(L, y, t')\right]^2 \right\}.
\end{aligned}
\end{equation}
% The third loss function term imposing the initial condition is given by $\mathcal{L}_{\mathrm{ic}}=\int_{-L}^Ld{x}\int_{-L}^Ld{y}\left[\phi\left(x,y,0\right)-\phi_0\right]^2$. The last loss function term is given by $\mathcal{L}_{\mathrm{con}}=\int_0^{t}d{t'}\int_{-L}^Ld{x}\int_{-L}^Ld{y} \left(\partial_t{\phi}+\nabla \cdot \mathbf{J}\right)^2$, which takes into account of the physical constraint of $\phi$ conservation, as given in Eq.~(\ref{Eq:App2-phit}). 
The initial condition is given by $\mathcal{L}_{\mathrm{ic}}=\int_{-L}^Ld{x}\int_{-L}^Ld{y}\left[\phi\left(x,y,0\right)-\phi_0\right]^2$, which is the third loss function term. The last loss function term is $\mathcal{L}_{\mathrm{con}}=\int_0^{t}d{t'}\int_{-L}^Ld{x}\int_{-L}^Ld{y} \left(\partial_t{\phi}+\nabla \cdot \mathbf{J}\right)^2$, which takes into account of the physical constraint of $\phi$ conservation, as given in Eq.~(\ref{Eq:App2-phit}). 

In Fig.~\ref{Fig:2dCH}, we compare the trained solutions $\phi(x,t)$ using the DOMM and the numerical solutions using the traditional FDM~\cite{GAO20121372} at several different times, respectively. From this figure, we can see that the DOMM can solve the two-phase dynamics in 2D with high accuracy and convergence. The trained solutions are very close to the numerical solutions with final MSE less than $10^{-5}$, implying high accuracy. Moreover, the errors and the loss function are both found to decay systematically with increasing training steps, ensuring training convergence. All the trained results are obtained using the same network structure with width $N_{\mathrm{w}}=128$, depth $N_{\mathrm{d}}=3$. We choose the weights in the loss function to be $w_{\mathrm{ic}} = 1000.0$ and $w_{\mathrm{om}} = w_{\mathrm{con}} = 10.0$ and $w_{\mathrm{bc}} = 1.0$. Moreover, we have also used the strategy of time splitting with time interval $\Delta t = 0.1$ to accelerate the training of the 1D dynamic process.

\section{Application 3: Solving two-phase hydrodynamics}\label{Sec:App3}

\begin{figure}[!ht]%
  \centering
\includegraphics[width = 0.9\linewidth]{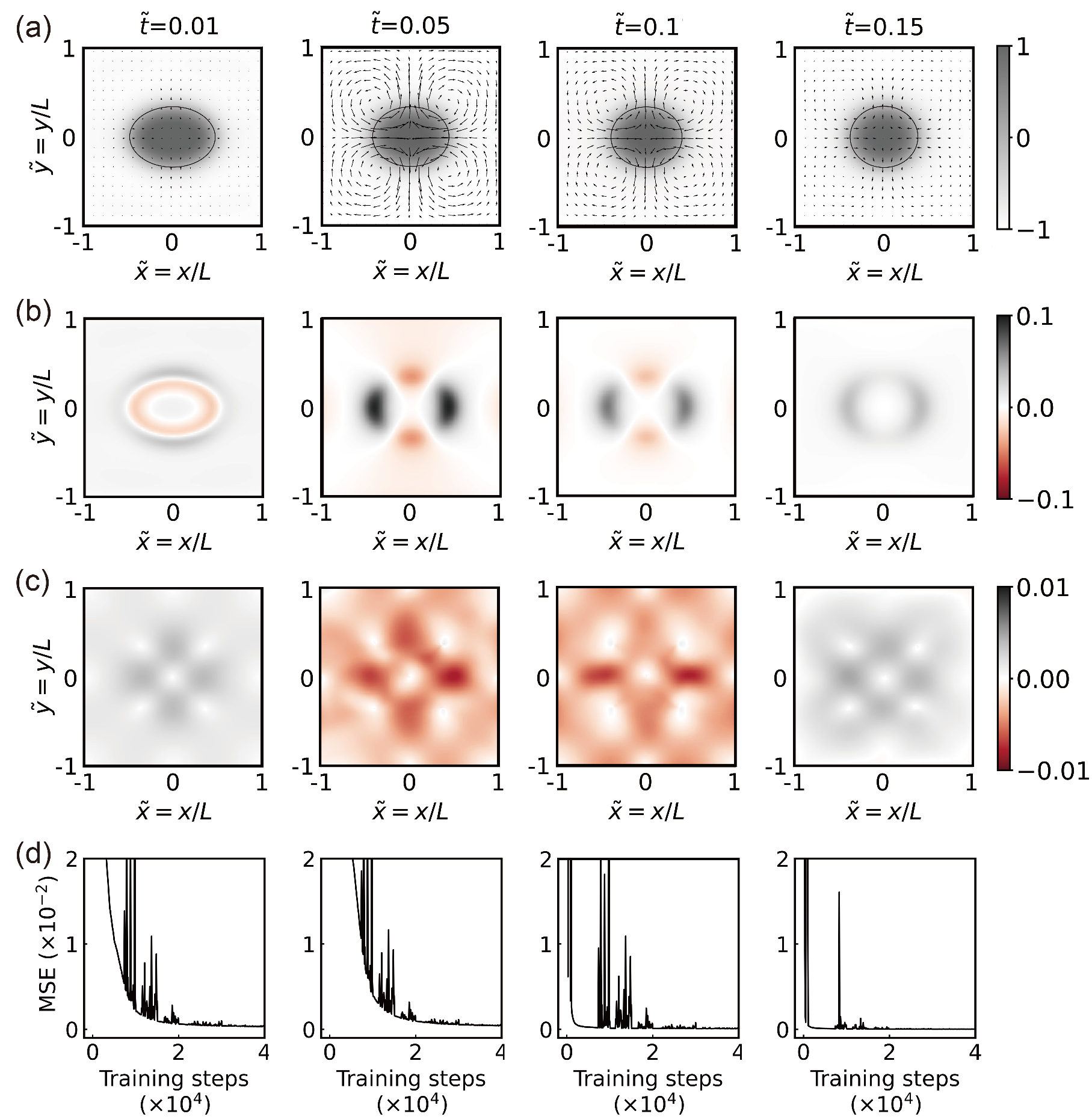}
\caption{Application 3 -- Solving 2D two-phase hydrodynamics: the relaxational dynamics of an initially elliptical droplet to its final equilibrium circular shape upon periodic boundary condition at $x,\, y=\pm L$. (a) Trained dynamical process of the droplet relaxation using DOMM including both phase dynamics (phase field, gray map) and hydrodynamics (flow field, arrows). The time $t$ is normalized by the unit $\tau\equiv L^2/Ma$, the interface thickness is taken to be $\epsilon = 0.05 L$, and the dimensionless parameter $\mathcal{B} = M\eta/L^2$ is taken to be $0.1$. (b) and (c) The pointwise error maps of the trained solution of $\phi$ and the speed (velocity magnitude) are calculated using $\left(\phi_{\mathrm{DOMM}}(\mathbf{x},t)-\phi_{\mathrm{FDM}}(\mathbf{x},t)\right)$ and $\left(v_{\mathrm{DOMM}}(\mathbf{x},t)-v_{\mathrm{FDM}}(\mathbf{x},t)\right)$, respectively, where $\phi_{\mathrm{DOMM}}$ and $v_{\mathrm{DOMM}}$ are the trained solutions by DOMM; $\phi_{\mathrm{FDM}}$ and $v_{\mathrm{FDM}}$ are the numerical solutions by the traditional FDM. 
(c) Convergence of training MSE of $\phi$ as a function of training steps at the time corresponding to that in (a). All trained results are obtained using the same neural network structure with width $N_{\mathrm{w}}=128$, depth $N_{\mathrm{d}}=5$.}
\label{Fig:2dSCH}
\end{figure}

In this section, we consider the application of DOMM to a more complicated dynamic process~\cite{Qian2006,zhao2021PINN} -- two-phase hydrodynamics at zero Reynolds number in 2D: the relaxational dynamics of an initially elliptical droplet to its final equilibrium circular shape as shown in Fig.~\ref{Fig:2dSCH}. In this case, two dynamic physics processes are closely coupled: the phase field dynamics and the viscous flow. As mentioned above, in dynamic processes involving multi-physics couplings, the DOMM gives physically meaningful weights to the loss function of different dynamic equations governing different physics processes, which decreases the number of adjustable weights and is regarded as one of the major advantages of DOMM over PINN where each dynamic equation is included into the loss function with an independent adjustable weight~\cite{zhao2021PINN}.

In the two-phase hydrodynamic process at zero Reynolds number in 2D, the state variable is $\phi(\mathbf{x},t)$ and velocity field $\mathbf{v}(\mathbf{x},t)=(v_x,v_y)$, the phase field parameter at position $\mathbf{x}=(x,y)$ and time $t$. Firstly, the flow is assumed to be incompressible, $\nabla \cdot \mathbf{v}=0$; the equations of $\phi$ conservation and force balances are given by
\begin{equation}\label{Eq:App3-phit}
\partial_t{\phi}+\nabla \cdot \left(\phi \mathbf{v}+\mathbf{J} \right)=0,
\end{equation}
\begin{equation}\label{Eq:App3-stress}
\nabla \cdot (\boldsymbol{\sigma}-\boldsymbol{\Pi})=0,
\end{equation}
where $\boldsymbol{\sigma}(\mathbf{x},t)$ and $\boldsymbol{\Pi}$ are the viscous stress tensor and reversible pressure tensor, respectively. The free energy $\mathcal{F} [\phi(\mathbf{x},t)]$ of the system is still given by Eq.~(\ref{Eq:App2-F}). The dissipation function $\Phi$ takes the following quadratic form
\begin{equation}\label{Eq:App3-Phi}
\Phi[\mathbf{J}(\mathbf{x},t),\boldsymbol{\sigma}(\mathbf{x},t)]=\int d\mathbf{x} \left( \frac{\mathbf{J}^{2}}{2M} + \frac{\boldsymbol{\sigma}^{2}}{4\eta}\right),
\end{equation}
with $M$ and $\eta$ being the constant mobility and viscosity, respectively. Note that here we use the viscous stress (representing momentum flux) being the viscous dissipation rate~\cite{deGennes1993} in contrast with most theoretical works that use the viscous dissipation in terms of quadratic shear rates~\cite{Qian2006,Doi2013}. Minimizing the Rayleighian $\mathcal{R}[\mathbf{J}(\mathbf{x},t)]=\dot{\mathcal{F}}+\Phi-\int d\mathbf{x} p\nabla \cdot \mathbf{v}$ with respect to the fluxes $\mathbf{J}$ and $\boldsymbol{\sigma}$ yields the most probable fluxes
\begin{equation}\label{Eq:App3-Jstar}
\mathbf{J}^{*}=-M\nabla \hat{\mu},
\end{equation}
\begin{equation}\label{Eq:App3-sigmastar}
\boldsymbol{\sigma}^{*}=\eta \left(\nabla \mathbf{v}+ \nabla \mathbf{v}^{\mathrm{T}}\right),
\end{equation} 
where $p$ is the Lagrange multiplier for the incompressibility constraint, the superscript $T$ denotes the tensor transpose, the generalized chemical potential is given in Eq.~(\ref{Eq:App2-mu}), and the reversible pressure tensor satisfies the identity $\nabla\cdot \boldsymbol{\Pi}= \nabla p + \phi \nabla \hat{\mu}$. Substituting Eqs.~(\ref{Eq:App3-Jstar}) and ~(\ref{Eq:App3-sigmastar}) into Eqs.~(\ref{Eq:App3-phit}) and (\ref{Eq:App3-stress}) give the standard Stokes-Cahn-Hilliard equation
\begin{equation}\label{Eq:App3-SCH-CH}
\partial_t{\phi}-M\nabla^2 \hat{\mu}=0,
\end{equation} 
\begin{equation}\label{Eq:App3-SCH-S}
-\nabla p + \eta \nabla^2\mathbf{v} + \hat{\mu}\nabla\phi=0,
\end{equation} 
supplemented with $\nabla \cdot \mathbf{v} = 0$, where we have rewritten the capillary force term $-\phi\nabla \hat{\mu}$ into the more widely used form $\hat{\mu}\nabla \phi$ by absorbing a term $-\phi \hat{\mu}$ into the pressure~\cite{Qian2006}. In addition, the Onsager-Machlup integral is given by
\begin{equation}\label{Eq:App3-OM}
% \mathcal{L}_{\mathrm{om}} = \int_0^{t}d{t} \left\{\mathcal{R}[j]-\mathcal{R}[j^*]\right\}= 
\mathcal{L}_{\mathrm{om}}= \int d{t}\int d\mathbf{x} \left[\frac{1}{2M}\left(\mathbf{J}-{\mathbf{J}^*}\right)^2+ \frac{1}{4\eta} \left(\boldsymbol{\sigma}-\boldsymbol{\sigma}^*\right)^2\right]. 
\end{equation} 
In the use of DOMM to solve the two-phase hydrodynamics in 2D, we take the periodic boundary conditions and the following initial condition
\begin{equation}\label{Eq:App2-2DSCH-ic}
\phi_0(x,y)=\tanh\left(0.5-\epsilon^{-1}\sqrt{x^2+(1.5y)^2} \right),
\end{equation}
for $-L\leqslant x,\,y \leqslant L$ and $t \geqslant 0$. 
The input parameters of the neural network are the position vector $\mathbf{x}=(x,y)$ and time $t$; the output parameters are chosen to be the phase-parameter field $\phi$, velocity field $\mathbf{v}$, the diffusion flux vector $\mathbf{J}$ and the viscous stress tensor $\boldsymbol{\sigma}$. The loss function minimized to train the neural network still takes the general form of Eq.~(\ref{Eq:DOMM-Loss}). Here the first loss function term $\mathcal{L}_{\mathrm{om}}$ is the Onsager-Machlup integral in Eq.~(\ref{Eq:App3-OM}) with $\mathbf{J}^*$ and $\boldsymbol{\sigma}^*$ given in Eqs.~(\ref{Eq:App3-Jstar}) and (\ref{Eq:App3-sigmastar}), respectively. The second loss function term is associated with the periodic boundary conditions, taking a similar form of Eq.~(\ref{Eq:App2-2DCH-LossBC}) for the periodic boundary conditions of $\phi$, $\partial_x\phi$, $\partial_y\phi$, $v_x$ and $v_y$. 
The third loss function term imposing the initial condition is given by $\mathcal{L}_{\mathrm{ic}}=\int d\mathbf{x}\left\{\left[\phi\left(\mathbf{x},0\right)-\phi_0\right]^2 + \mathbf{v}^2\left(\mathbf{x},0\right)/V_0^2 \right\}$ with $V_0\equiv L/\tau$ and $\tau\equiv L^2/Ma$ being the units of velocity $\mathbf{v}$ and time $t$, respectively. The last loss function term is given by $\mathcal{L}_{\mathrm{con}}=\int_0^{t}d{t'}\int d\mathbf{x} \left\{ \left( \nabla \cdot \boldsymbol{\sigma} + \mu\nabla\phi\right)^2 + \left[\partial_t{\phi}+\nabla \cdot \left(\phi\mathbf{v}+ \mathbf{J}\right)\right]^2 + \left( \nabla \cdot \mathbf{v}\right)^2 \right\}$, which takes into account of the physical constraints of $\phi$ conservation in Eq.~(\ref{Eq:App3-phit}) and the incompressibility condition $\nabla \cdot \mathbf{v}=0$. 

In Fig.~\ref{Fig:2dSCH}, we compare the trained solutions $\phi(\mathbf{x},t)$ and $\mathbf{v}(\mathbf{x},t)$ using the DOMM and the numerical solutions using the traditional FDM~\cite{GAO20121372} at several different times, respectively. From this figure, we can see that the DOMM can solve the two-phase hydrodynamics in 2D that involves multi-physics coupling with good accuracy and convergence. The trained solutions are very close to the numerical solutions with final MSE less than $10^{-3}$, implying good accuracy. Moreover, the errors and the loss function are both found to decay systematically with increasing training steps, ensuring training convergence. All the trained results are obtained using the same network structure with width $N_{\mathrm{w}}=128$, depth $N_{\mathrm{d}}=5$. We choose the weights in the loss function to be $w_{\mathrm{ic}} = 1000.0$ and $w_{\mathrm{om}} = w_{\mathrm{con}} = 10.0$ and $w_{\mathrm{bc}} = 1.0$. Moreover, we have also used the sampling strategy of time splitting with time interval $\Delta t = 0.1$ to accelerate the training of the 2D dynamic process.

\section{Concluding remarks}\label{Sec:Conclusions}

In this work, we have proposed a unified computational method for soft matter dynamics -- the deep Onsager-Machlup method (DOMM), which combines the brute forces of deep neural networks (DNNs) with the fundamental physics principle -- Onsager-Machlup variational principle (OMVP). In the DOMM, the trial solution to the dynamics is constructed by DNNs that allow us to explore a rich and complex set of admissible functions. It outperforms the representation using Doi's Ritz-type method where one has to impose carefully-chosen trial functions~\cite{Doi2015,Doi2019,Doi2021}. This endows the DOMM with the potential to solve rather complex problems in soft matter dynamics that involve multiple physics with multiple slow variables, multiple scales, and multiple dissipative processes.
% DOMM has the capacity of mining high-dimensional information from deep neural networks.
In comparison to traditional numerical methods such as FEM and FDM, machine learning (ML) methods are less competitive in solving low-dimensional problems, where many efficient and accurate traditional methods have already been well developed. However, ML methods including DOMM demonstrate several advantages over traditional methods in some other scenarios. For example, ML methods are mesh-free methods, capable of avoiding discretization on grid points, which makes them efficient for high-dimensional problems and/or with complex boundary geometries. Moreover, ML methods are interpolation-free methods, allowing to calculate solutions at spatiotemporal points not included in the sampled computational set but within the computational domain without interpolation. This calculation process is efficient and fast (simply input the desired spatiotemporal points into the neural network). In addition, ML methods do not impose restrictions on the time variable; one can deal with time in a similar way as spatial variables. In contrast, traditional numerical methods usually set a strict constraint between the time step size and the spatial resolution to ensure convergence. Moreover, in comparison to other machine learning methods proposed particularly for soft matter physics~\cite{Perdikaris2019,Faroughi2022}, the DOMM has several advantages as follows. It is naturally nonlinear, naturally adaptive, and relatively insensitive to the complexity of soft matter dynamics, of the OM functional, of the dimensions of both physical space and state variables as well as of the order of the derivatives of state variables. 
Here the DOMM has been used to solve several typical dynamic problems in soft matter physics: particle diffusion in dilute solutions, two-phase dynamics with and without hydrodynamics. The predicted results from DOMM agree very well with the exact solution or numerical solution from the traditional finite difference method (FDM). These results validate the DOMM and justify its accuracy and convergence as a powerful alternative computational method in solving soft matter dynamics. 

Below we make a few general remarks and outlook.

(i) We observed that the solution training using DOMM follows the frequency principle found by Xu \emph{et al}~\cite{Xu_2020}, which states that a DNN tends to learn a target function from low to high frequencies during the training. For diffusion dynamics and 2D two-hase hydrodynamics, the solutions in the latter time were well-trained first because the solutions get smoother (with lower frequencies or smaller wavenumbers in physics language) with increasing time. In contrast, for the 1D artificially-designed two-phase dynamics studied in Sec.~\ref{Sec:App2}, the solutions in the earlier time were well-trained first because the solutions get sharper (with higher frequencies or larger wavenumbers in physics language) with increasing time. Therefore, we expect that for other dynamic processes such as Burgers equation with shock formation where the solution gets sharper (with larger frequencies) with time, the frequency principle should also work and one would observe that the solutions in the earlier time were well-trained first. We should carry out such studies in the future.

(ii) We should compare with PINN and its variants~\cite{Perdikaris2019,Karniadakis2021physics} further to solve the complex soft matter dynamics derived where multiple physics are closely coupled and DOMM includes physically-meaningful weights in the OM integral. 

Particularly, PINNs are known to be able to study not only forward problems of solving partial differential equations (PDEs) but also show unique advantages in solving inverse problems~\cite{Bao_2020,Perdikaris2019,Karniadakis2021physics} where some information in the PDEs is not complete, for example, initial or boundary data, physical coefficients or applied forces are unknown. In the future, we should also explore the applications of DOMM to study inverse problems by calibration of variational models of soft matter dynamics to observable field quantities obtained from experiments or numerical computations. 

(iii) The OMVP and the minimization of OM integral can also be used to study rare events~\cite{Weinan2004,Touchette2009,Ren2020} that occur with low frequency but have a widespread effect and might destabilize systems, for example, nucleation events during phase transitions, conformational changes in macromolecules, and chemical reactions. Therefore, one of our major future directions to extend the applications of DOMM is to study rare events in soft matter physics. 

(iv) Many transport equations in soft matter dynamics can be derived from OVP and OMVP~\cite{Doi2013,Doi2021,Xu2017}. However, in this work, we have only considered the applications of DOMM to solve some of the simplest soft matter dynamics: diffusion and two-phase dynamics. DOMM could be capable of computing more complex soft matter dynamics~\cite{Doi2013,Doi2021,Xu2017,Xu2021}, such as spinodal decomposition in solutions, elasto-diffusion in gels, nemato-hydrodynamics in liquid crystals, and active nemato-hydrodynamics in active nematics, \emph{etc}. 

(v) The OMVP is a general principle for non-equilibrium statistical thermodynamics~\cite{Onsager1953} and has wide applications in soft matter dynamics~\cite{Doi2019, Doi2021}. It has recently been extended by us to the study of active matter physics~\cite{Xu2021}. Particularly, in an active (self-propelled) particle system, the state variables $\boldsymbol{\alpha}$ represent the spatial coordinates of each particle. In this case, the DOMM method can be used to study the rare events~\cite{Avishek2022} in the active matter and inverse problems, for example, learn the potential landscape from the stochastic trajectories of active particles~\cite{genkin2021learning}.
\section*{Acknowledgments}
% The author would like to thank  ....

The authors thank Tiezheng Qian, Zhen Zhang, Masao Doi, Penger Tong, and Shigeyuki Komura for useful discussions. X. Xu is supported in part by National Natural Science Foundation of China (NSFC, No.~12374209, No.~12004082, No.~12131010). D. Wang is partially supported by National Natural Science Foundation of China (Grant No. 12101524, 12422116), Guangdong Basic and Applied Basic Research Foundation (Grant No. 2023A1515012199), Shenzhen Science and Technology Innovation Program (Grant No. JCYJ20220530143803007, RCYX20221008092843046, GXWD20201231105722002-202008291\\
62111001), Guangdong Provincial Key Laboratory of Mathematical Foundations for Artificial Intelligence (2023B1212010001), and Hetao Shenzhen-Hong Kong Science and Technology Innovation Cooperation Zone Project (No.HZQSWS-KCCYB-2024016).

%%%% Bibliography  %%%%%%%%%%

% \begin{thebibliography}{99}
% \bibitem{Berger}M. J. Berger and P. Collela, Local adaptive mesh refinement
% for shock hydrodynamics,
% J. Comput. Phys., 82 (1989), 62-84.
% \bibitem{deBoor}C. de Boor,  Good Approximation By Splines With Variable Knots II, in Springer Lecture
%  Notes Series 363, Springer-Verlag, Berlin, 1973.
% \bibitem{TanTZ} Z. J. Tan, T. Tang and Z. R. Zhang, A simple moving mesh method for one- and
% two-dimensional phase-field equations, J. Comput. Appl. Math., to appear.
% \bibitem{Toro}E. F. Toro, Riemann Solvers and Numerical Methods for Fluid Dynamics,
% Springer-Verlag Berlin Heidelbert, 1999.

% \end{thebibliography}
\bibliographystyle{unsrt}
\bibliography{rsc}

\end{document}